\documentclass[fleqn,usenatbib]{mnras}

\usepackage{newtxtext,newtxmath}

\usepackage[T1]{fontenc}

\DeclareRobustCommand{\VAN}[3]{#2}
\let\VANthebibliography\thebibliography
\def\thebibliography{\DeclareRobustCommand{\VAN}[3]{##3}\VANthebibliography}

\usepackage{graphicx}	
\usepackage{amsmath}	
\usepackage{url}

\newcommand{\msol} {M$_{\odot}$}

\newcommand{\about} {$\sim$}

\title[GP of Type II and IIb SN lightcurves]{What can Gaussian Processes really tell us about supernova lightcurves? Consequences for Type II(b) morphologies and genealogies}

\author[H. F. Stevance et al.]{
H. F. Stevance,$^{1,2}$\thanks{E-mail: hfstevance@gmail.com}
A. Lee,$^{1}$
\\
$^{1}$Department of Physics, The University of Auckland, Private Bag 92019, Auckland, New Zealand\\
$^{2}$Astrophysics Research Centre, School of Mathematics and Physics, Queen’s
University Belfast, N. Ireland, BT7 1NN, United Kingdom\\
}
\date{Accepted XXX. Received YYY; in original form ZZZ}

\pubyear{2015}

\begin{document}
\defcitealias{pessi2019}{P19}
\defcitealias{gutierrez2017}{G17}
\label{firstpage}
\pagerange{\pageref{firstpage}--\pageref{lastpage}}
\maketitle

\begin{abstract}
Machine learning has become widely used in astronomy. Gaussian Process (GP) regression in particular has been employed a number of times to fit or re-sample supernova (SN) light-curves, however by their nature typical GP models are not suited to fit SN photometric data and they will be prone to over-fitting.
Recently GP re-sampling was used in the context of studying the morphologies of type II and IIb SNe and they were found to be clearly distinct with respect to four parameters:  the rise time (t$_{\rm rise}$), the magnitude difference between 40 and 30 days post explosion ($\Delta m_{\rm 40-30}$), the earliest maximum (post-peak) of the first derivative (dm1) and  minimum of the second derivative (dm2).
Here we take a close look at GP regression and its limitations in the context of SN light-curves in general, and we also discuss the uncertainties on these specific parameters, finding that dm1 and dm2 cannot give reliable astrophysical information.
We do reproduce the clustering in t$_{\rm rise}$--$\Delta m_{\rm 40-30}$ space although it is not as clear cut as previously presented.
The best strategy to  accurately populate the t$_{\rm rise}$-- $\Delta m_{\rm 40-30}$ space will be to use an expanded sample of high quality light-curves (such as those in the ATLAS transient survey) and analytical fitting methods.
Finally, using the BPASS fiducial models, we predict that future photometric studies will reveal clear clustering of the type IIb and II light curve morphologies with a distinct continuum of transitional events.  
\end{abstract}

\begin{keywords}
methods: statistical -- supernovae: general -- stars: evolution
\end{keywords}



\section{Introduction}
Massive stars, born with more than \about 8\msol, will typically end their lives as core-collapse supernovae (CCSNe).
The chemical composition of the envelope of the progenitor star will strongly influence the observed properties of the explosion, and before the physical cause of these characteristics were known, astronomers began using them to separate SNe into distinct categories.
The first divide split SNe with (type II) and without (type I) hydrogen signatures in their spectra, and despite the ``provisional" nature of the classification at the time, it endured \citep{minkowski1941}.
Further subdivision in the hydrogen-rich types was established on the basis of the shape of their light curves: type IIP for those showing a plateau, type IIL for those showing a linear decline.
Sub-categories of hydrogen-poor SNe were also created according to the presence (Ib) or absence (Ic) of helium in their spectra (for a review of SN types see \citealt{filippenko1997}).

The divide between the classes however is not a sharp one: some type II supernovae became type Ib later in life and were therefore called type IIb  \citep{filipenko1988}, suggesting that their progenitors were stripped of most, yet not all, of their hydrogen, retaining \about 0.5\msol \citep{smith2011}.
Additionally, in depth studies of the growing sample of type II SNe performed in the last decade have revealed a correlation between the amount of hydrogen in the ejecta of these SNe and the decay rate of their light curves \citep{anderson2014, sanders2015}. 
These observations point to a SN spectrum rather than a SN classification, Type IIP $\rightarrow$ Type IIL $\rightarrow$ Type IIb $\rightarrow$ Ib/c,  where the deciding factor in the resulting classification is the level of stripping of the hydrogen envelope of the progenitor star.

Some studies however have highlighted sharper divides between type II and type IIb SNe.
In analysing type II SN light curves \cite{arcavi2012} found that type IIb consistently showed more rapid decline rates, however the sample used at the time was rather small and there are now published examples of fast-declining type II SNe (e.g. SN2013ai \citealt{davis2021}).
More recently \cite{pessi2019} -- hereafter \citetalias{pessi2019} -- used innovative data analysis methods to create parameter spaces where type II and IIb light curves could be compared. 
They define the following features (for further detail see their section 4): 
\begin{itemize}
    \item $t_{\rm rise}$: time between the explosion date and maximum light.
    \item $\Delta m_{\rm 40-30}$: Magnitude difference between the phase 30 and phase 40 post explosion.
    \item dm1: The earliest maximum (after maximum light) of the first derivative of the light curve.
    \item dm2: The earliest minimum (after maximum light) of the second derivative of the light curve.
\end{itemize}

The resulting comparison yielded a very sharp divide between type II and IIb SNe, which they interpreted as potentially indicating that the evolutionary channels of type II and IIb SNe are separate. 

One cause for concern however is the use of Gaussian processes (GP) to re-sample their SN light curves before evaluating the features. 
GP are a supervised machine method that can be used to perform regression on data sets \citep{rasmussenw06} - the trained algorithm can then be used to interpolate between the data points.
Although they have been used with some success in fitting type Ia light-curves \citep{kim2013} and as a means to augment photometric data to create photometric classifiers \citep{boone2019}, GPs are by their nature not well suited to fitting SN light-curves as they assume unvarying timescale for the evolution of the function being fitted (see Section \ref{sec:length_scale} for an extended discussion). 
This is not consistent with the known physics of SNe.
Parametric methods such as Bazin fits \citep{bazin2009, villar2019} or the fits performed by \cite{sanders2015} (see their section 3) are physics-informed even if they are not physical simulations: they are created to be able to reproduce known characteristics of SN II light curves such as their plateaux and exponential decays. 
GP on the other hand may have a tendency to over-fit the data.

Therefore in this paper we use a restricted sample of type II and IIb SNe (see Section \ref{sec:thesample}) to offer an extended discussion of what GPs are, what can be achieved and where they fall short in the context of SN light-curve fitting (Section \ref{sec:gp}). 
We then look at the parameter space defined by \citetalias{pessi2019} and comment on which parameters are reliable (Section \ref{sec:results}), before discussing the implications and expectations of future studies of type II(b) light curve morphologies in the context of theoretical models of SN progenitors in the Binary Population and Spectral Synthesis (BPASS) code (\citealt{eldridge2017,stanway2018} -- Section \ref{sec:discussion}).
Finally we summarise and conclude in Section \ref{sec:conclusion}. 

\section{The sample}
\label{sec:thesample}
Because one of our main goals is to take a detailed look at the limitations of GPs and what they can tell us about the morphology of type II SNe and their progenitors, we create a restricted sample containing only high quality data.
There are two main criteria we select for: The data must be evenly sampled in time (avoid large gaps in the light curve) and it must have a well studied explosion dates.
This will give our GP models the best chance to perform well, although consequences of uneven sampling are already visible in some of the SNe we selected (e.g. SN2003hn see Section \ref{sec:results}) -- we chose not to remove them from the sample \textit{a posteriori}.
The smaller sample size also has the added advantage that each object and its fits could be reviewed very thoroughly. 
We make our full sample, as well as the codes used for the analysis available on GitHub (see Data Availability) 

We gather the light curves using the API of the Open Supernova Catalogue \citep{guillochon2017}\footnote{Note that the front-end of the catalogue is no longer available but the database remains available on GitHub. Unfortunately it is no longer being updated as of April 8th, 2022.} and most of the explosion dates from \cite{gutierrez2017} (hereafter \citetalias{gutierrez2017}) and \citetalias{pessi2019}. 
The light curves were visually inspected to select those which best fit our quality criteria. 
Some of the late time data were cropped to avoid gaps in the light curve and because the parameters defined by \citetalias{pessi2019} do not need call for data beyond 80 days.
We do none the less include data beyond 80 days as it is relevant to the broader discussion of fitting type IIP SN light curves with GP.

In total we selected 8 type IIb (Figure \ref{fig:LC_IIb}) and 14 type II (Figure \ref{fig:LC_II}) SNe which are listed in Table \ref{tab:sn_sample} including the time crop we applied and the explosion dates we will be using in this study.
For most SNe there are two explosion date estimates: one from using the Supernova Identification software (SNID; \citealt{blondin2007}), which uses spectral template matching to derive explosion dates based on a library of other SNe,  and one from pre-explosion non-detections, where the explosion date is taken as the mid-point between the last non-detection and the first observation.
The explosion data we used for each target is summarised in Table \ref{tab:sn_sample}.

In most cases both estimate were equivalent and we use the data derived from observations (marked as [obs.]). 
When the two estimates differed slightly we used the average between the two estimates ([avg.])
In a few cases, however, the SNID estimate seemed more appropriate, such as SN 2013df (the observationally derived explosion date would imply a shock-break-out and cooling tail lasting about two weeks which is inconsistent with expectations for type IIb SNe).
In the case of SN 2006el, both dates seem to intersect with the light curve, which is non-physical.
In this case we made use of the 5-day intrinsic error on the spectral fitting done by SNID quoted in \citetalias{pessi2019} and took the explosion date to be their SNID estimate -- 5 days. 
A similar situation arose with SN 2014cx, and the same solution is employed. 
Finally, in the case of SN 2008aq the SNID estimate is much lower than that from observations, but since the explosion epoch given by \cite{stevance2016} is much more consistent with the estimate from pre-explosion images, we do not take the average with the SNID values as we are concerned this would lead to an underestimate of the explosion date; we therefore use the literature value from \cite{stevance2016}.

\begin{table*}
	\centering
	\caption{Sample of SNe used in this study split by type and then in descending order of data quality (visual examination). The V-band light curve data were downloaded from the Open Source Supernova catalogue but not all downloaded data were included (e.g. spurious or noisy measurements were removed). In column 2 we show the references of for the data used in the final sample.  \textbf{Keys:} T.1[obs.] -- explosion date from table 1 `Observed' column | T.1[SNID] -- explosion date from table 1, `SNID' column | T.1[avg.] -- explosion date is the average of the `Observed' and `SNID' columns | T.A1 -- explosion date from table 1 in appendix A \citetalias{pessi2019} | T.A2[obs.] -- explosion date from table 2 in appendix A of \citetalias{gutierrez2017} column 7 | T.A2[SNID] -- explosion date from table 2 in appendix A column 10  | T.A2[avg.] -- explosion date from table 2 in appendix A of  \citetalias{gutierrez2017}; average between column 7 and 10. For more details on the explosion dates of SN 2006el, SN 2008aq and SN 2014cx see main text. }
	\label{tab:sn_sample}
	\begin{tabular}{p{2cm}p{5cm}p{2cm}p{1.6cm}p{2.5cm}}
		\hline
		Name & Open Supernova Catalogue photometry reference(s) & Dates included & Exp. date used [MJD]  & Exp. date reference\\
		\hline
		\textbf{Type IIb} \\
		SN 2008ax & \cite{pastorello2008, tsvetkov2009} & < 54,700 [MJD] & 54528.3 & T.1[obs.] \citetalias{pessi2019}\\
		SN 2011dh & \cite{brown2014} & all & 55712.4 & T.1[obs.] \citetalias{pessi2019}\\
		SN 1993J & \cite{okyudo1993,vandriel1993,benson1994,barbon1995,richmond1996} & < 49,175 [MJD] & 49073.1 & T.1[avg.] \citetalias{pessi2019}\\
		SN 2006T & \cite{bianco2014,stritzinger2018} & all & 53759.0 &  T.1[obs.] \citetalias{pessi2019}\\
		SN 2004ex & \cite{stritzinger2018} & all & 53287.9 & T.1[avg.] \citetalias{pessi2019}\\
		SN 2008aq & \cite{brown2014} & all & 54513.0 & \cite{stevance2016}\\
		SN 2013df & \cite{morales-Garoffolo2014,brown2014} & < 56,550 [MJD] & 56448.2 & T.1[SNID] \citetalias{pessi2019}\\
		SN 2006el & \cite{drout2011,bianco2014} & all & 53959 & T.1[SNID - error] \citetalias{pessi2019}\\
		 --  \\
		\textbf{Type II}  \\
        SN2013ej & \cite{brown2014,huang2015,yuan2016,deJaeger2019} & < 56,700 [MJD] & 56497.5 & T.A1 \citetalias{pessi2019} \\
        SN2012aw &  \cite{brown2014,dallora2014,deJaeger2019} & < 56,200 [MJD] & 56002.0 &  T.A2[SNID] \citetalias{gutierrez2017}\\
        SN1999em &  \cite{elmhamdi2003,anderson2014,faran2014,galbany2016} & < 51,700 [MJD] & 51476.5 &   T.A2[obs.] \citetalias{gutierrez2017}\\
        SN2014cx &  \cite{brown2014,valenti2016} & < 57,100 [MJD] & 56897.9 &  T.A1 \citetalias{pessi2019} - 5 days \\
        SN2004er & \cite{anderson2014} & all &  53271.8 & T.A2[obs.] \citetalias{gutierrez2017} \\
        SN2009ib & \cite{takats2015} & < 55,250 [MJD] & 55045.0 & T.A1 \citetalias{pessi2019}\\
        SN2013fs &  \cite{valenti2016} & < 56,650 [MJD] & 56570.8 &  T.A1 \citetalias{pessi2019}\\
        SN2009N  &  \cite{takatas2014,brown2014,anderson2014,hicken2017,deJaeger2019} & all & 54845.4 & T.A2[col.10] \citetalias{gutierrez2017} \\
        SN2014G  &  \cite{deJaeger2019} & < 56,750 [MJD] & 56669.6 & T.A1 \citetalias{pessi2019}\\
        SN2003hn &  \cite{galbany2016} & all & 52866.5 & T.A2[obs.] \citetalias{gutierrez2017} \\
        SN2008aw & \cite{anderson2014} & < 54,600 [MJD] & 54520.3 & T.A2[obs.] \citetalias{gutierrez2017} \\
        SN1999gi & \cite{leonard2002} & all & 51518.2 & T.A1 \citetalias{pessi2019}\\
        SN1992ba &  \cite{anderson2014,galbany2016} & all & 48884.9 & T.A2[SNID] \citetalias{gutierrez2017} \\
        SN2013ai & \cite{valenti2016} & all & 56339.7 &  T.A1 \citetalias{pessi2019}\\
		\hline
	\end{tabular}
\end{table*}

\section{Gaussian Process regression}
\label{sec:gp}
\subsection{A Brief introduction to Gaussian Processes}
GP are at the heart of this study, so we provide a brief introduction to the method that should be accessible to the broader SN community (also see appendix A of \citealt{boone2019}). For an extended introduction aimed at astronomers we suggest consulting Chapter 3.4 of \cite{mcallister2017} and for an in depth statistical view we refer the reader to the textbook by \cite{rasmussenw06}\footnote{Official PDF of the textbook is freely available at \url{http://gaussianprocess.org/gpml/chapters/RW.pdf}}.

GP are a powerful machine learning method which can approximate a function of unknown form with a multivariate Gaussian distribution. 
In the context of fitting SN light-curves, the function underlying our data has one independent variable (time), and one dependent variable (magnitude).
Then the approximation of the observed magnitudes ($\textbf{m}$) given our observation dates ($\textbf{t}$\footnote{These variables are vectors each containing all the values for time and magnitude in the data}) made by the GP model can be written with the probability distribution function:
\begin{equation}\label{eq:GP1}
    P(\textbf{m}|\textbf{t}) = \mathcal{N}(\mu(\textbf{t}), \textbf{K})
\end{equation}
where $\mu(\textbf{t})$ is the mean function, which describes how the expected value of \textbf{m} varies as a function of \textbf{t}, and \textbf{K} is the covariance matrix. 

In essence, each data point is its own Gaussian, and training the GP model is understanding how they all relate to each other by optimizing $\mu(\textbf{t})$ and $\textbf{K}$ to best fit the data.
Subsequently the optimised model of Eq. \ref{eq:GP1} can be used to infer new magnitude values for our choice of dates.
This makes GP a powerful fitting and re-sampling method on two accounts: firstly we do not need a physical or analytical model; secondly it naturally provides uncertainties on the inferred data points.
A number of implementations exist which make the use of GP very accessible and easy to use. Two of the most popular are the GPy (GP specific) and the the {\tt sklearn} (general machine learning) python packages \citep{gpyopt2016,scikit-learn}. 

When training the GP algorithm, the most important consideration of the user is the choice of the \textit{covariance kernel}. 
There are a number of commonly used kernels with a variety of properties (for an in depth introduction see chapter 2 of \cite{duvenaud2014}\footnote{For an informal summary of this chapter consults \url{https://www.cs.toronto.edu/~duvenaud/cookbook/} }); the essential take away here is that the kernel is a function chosen to reflect characteristics expected to be seen in the functions we are approximating: e.g. an eclipsing binary light curves would be fit with a periodic kernel.

When fitting SN light curves the preferred kernels in the literature have been the squared-exponential kernel \citep{kim2013,fakhouri2015,pessi2019} and the Mat{\`e}rn--3/2 kernel \citep{boone2019}.
The former is the prototypical kernel used in many non-periodic applications of GP modelling; the latter results in much sharper variations in the model, which could be more suitable for type II plateau SNe light curves where their sharp drop off the plateau could be smeared by the squared-exponential kernel. 
\cite{boone2019} mention that they find little difference in fits performed with the Mat{\`e}rn--3/2 and the squared-exponential kernel, but we do not find that to be the consistently the case (see Section \ref{sec:matern}). 

In a first instance, we will use the squared-exponential kernel to perform our fits.
It is characterised by two hyper-parameters: the length scale and the variance. 
In theory, and crudely speaking, a smaller length scale results in a mean function $\mu(\textbf{t})$ with more frequent "wiggles" and a larger variance allows the data to deviate more from its mean (potentially increasing the amplitude of said "wiggles").
Depending on the GP implementation used and the data, the starting values chosen by the user may have a dramatic impact on the final fit: in practice we found that the  optimizers in GPy converged on the same solution when given a large range of input length scale and variance, whereas with {\tt sklearn} fits required more careful consideration of the boundary values for the length scale (see Section \ref{sec:sklearn}).

Another type of kernels used by \citetalias{pessi2019} were those meant not to reproduce the physical patterns in the data but the stochastic ones:
White noise or bias kernels can be added to other kernels when building the GP model. A white noise kernel is one which adds variance to model data without changing the mean function, whereas the bias kernel can also affect the meant function $\mu(\textbf{t})$.
As we will see, whether a noise kernel is included and how it is implemented can have an influence on the resulting model and whether it fits (or over-fits) the data and there are other avenues to taking into account noise that may not be taken into account in the error bars (see below the case of SN~2004ex).



\subsection{Kernel choice and kernel `engineering'}
\cite{boone2019} in their section 5.5 mention that further exploration of which kernels are most suitable to SN light curve fitting should be conducted. 
The choice of kernel is indeed the most important consideration in the construction of a GP model and it is not uncommon for different kernels to be combined in order to better reflect the behaviour of a data set -- this is sometimes called kernel `engineering'.
It is beyond the scope of this paper to systematically test how specific kernel combinations behave on a large sample of SN data, however we found that even within our constrained set of light-curves distinct kernel choices, and at times some basic kernel engineering, allowed for better fits.

\subsubsection{RBF and RBF+Bias}
\begin{figure*}
	\includegraphics[width=14cm]{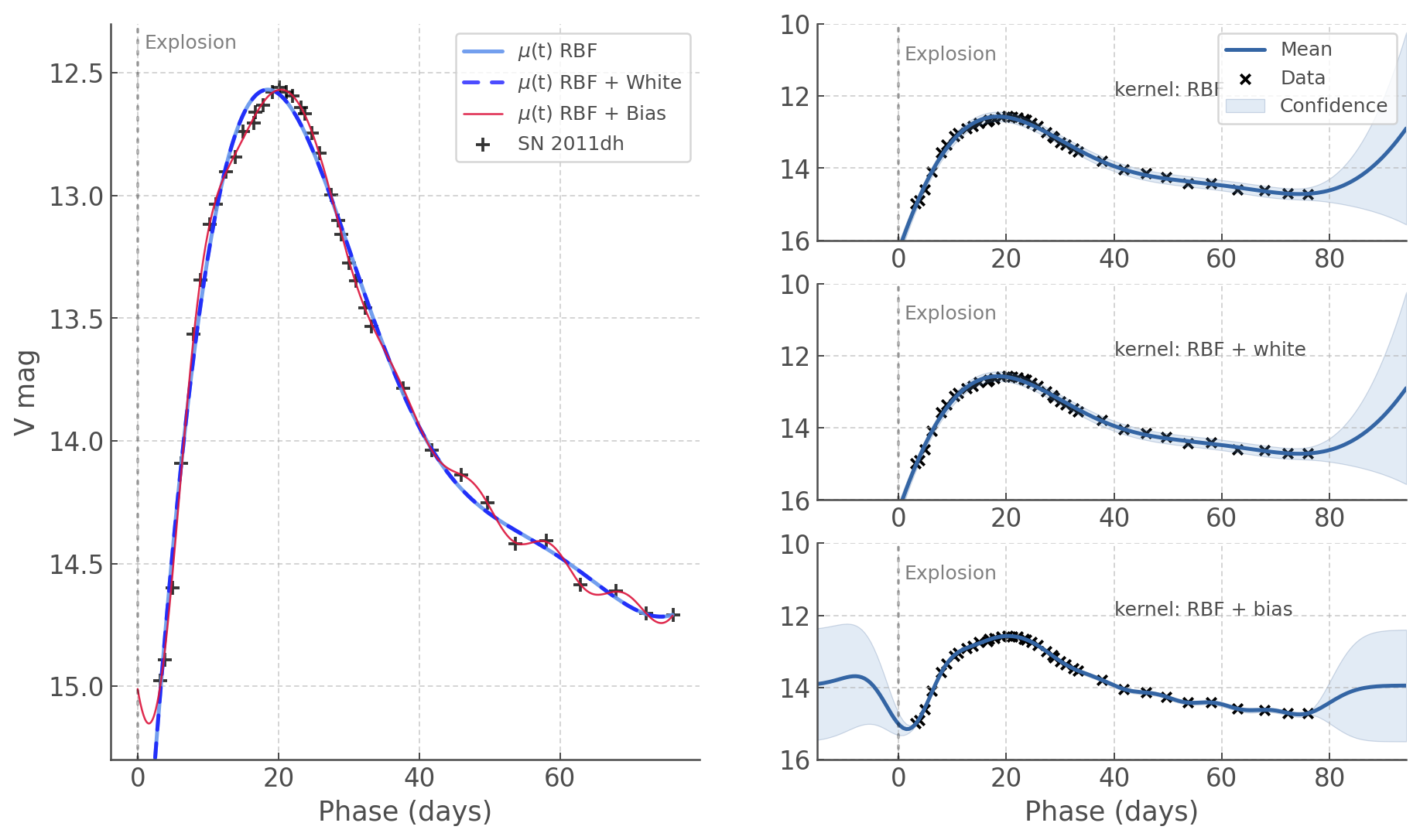}
    \caption{GPy fits of SN~2011dh using three different kernels. Note that on the first and second panels the variances on the fits are not shown to make the variations in the means easier to see.}
    \label{fig:08ax_11dh}
\end{figure*}

Here we present how results can be effected by including a noise kernels alongside a squared-exponential kernel, also known as the Radial Basis Function (RBF) kernel -- this configuration is what was used by \citetalias{pessi2019}.
We focus on SN~2011dh as an example (see Figure \ref{fig:08ax_11dh}) but note that other SNe show similar behaviour (e.g. SN~2008ax). 
Following \citetalias{pessi2019} the prior length scale and variance used for the RBF kernel are the average time interval between observed epochs and the standard deviation of that difference; the noise kernels were set up using the dot product of the errors on the magnitude as their variance.
Note that the results presented in this section were obtained with the GPy implementation and as mentioned above, these starting values have actually little bearing on the final model when the data is high quality and the kernel appropriate: in the case of SN~2011dh the models converge to a length scale of \about 35.5 when giving a starting length scales of 1000 or 0.01.
This is not true of fits produced with {\tt sklearn} see Section \ref{sec:sklearn}.

We can see in Figure \ref{fig:08ax_11dh} that, as expected, the RBF solutions and the RBF+white noise solutions for $\mu(\textbf{t})$  are identical, but including the bias kernel can have both positive and negative effects on the fit. 
The late time light curve is over fit by the RBF+bias kernel (it is also slightly over fit by the RBF kernel but not to the same degree). 
At the same time, the asymmetric peak of SN~2011dh is better reproduced by the RBF+bias kernel whilst the RBF kernel alone smooths it out, resulting in a slightly underestimated time at maximum (still within a couple of days). 

This leads to an interesting question: how are we to judge goodness of fit? 
Should one use a solution which can better reproduce the patterns of the early light curve at the risk of over-fitting the late times, or should one prefer smoother fit that risks under estimating the maximum light by a few days?
The answer obviously depends on the application of the fit -- if the purpose is to create model SN light curves over the full light-curve of the object over-fitting would be highly problematic whilst a systematic error on the maximum light could be relatively easy to account for. 
In this specific case we opted for the model with less over-fitting. 
Generally these considerations should be made transparent as the choices made by one team may not be applicable to another. 

Additionally we note that the goodness of fit was judged through visual examination. Tools such as least-squares or $\chi_2$ statistics are not useful here as a model that over-fits will perform best. Generally speaking methods of regularisation do exist to allow best-fit statistics to account for model complexity and discriminate against over-fitting, but the issue is that these methods are pure mathematical applications that do not take into account the known physics or behaviour of these time series, and what “over-fit” or “under-fit” is (not) acceptable. This is one of those few occasions where human expertise is hard to replace and for the visual examination we use two key criteria: Can the model fit the peak? Does it over-fit the linear decline/plateau?

We also use the Figure \ref{fig:08ax_11dh} to re-emphasise a typical word of caution when using machine learning methods: it is not appropriate to use a trained model outside the range of data it was trained on. 
When using the RBF kernel in GP, the length scale is typically the limit of reasonable extrapolation, however in the case of our SN light-curves it is clear that well within one length scale of the data the model already deviates form physical expectations. 
Consequently GP can be powerful interpolation tools, but similarly to splines they cannot reliably be used for extrapolation in the same way a physically motivated analytical model would.

\begin{figure}
    \includegraphics[width=7.5cm]{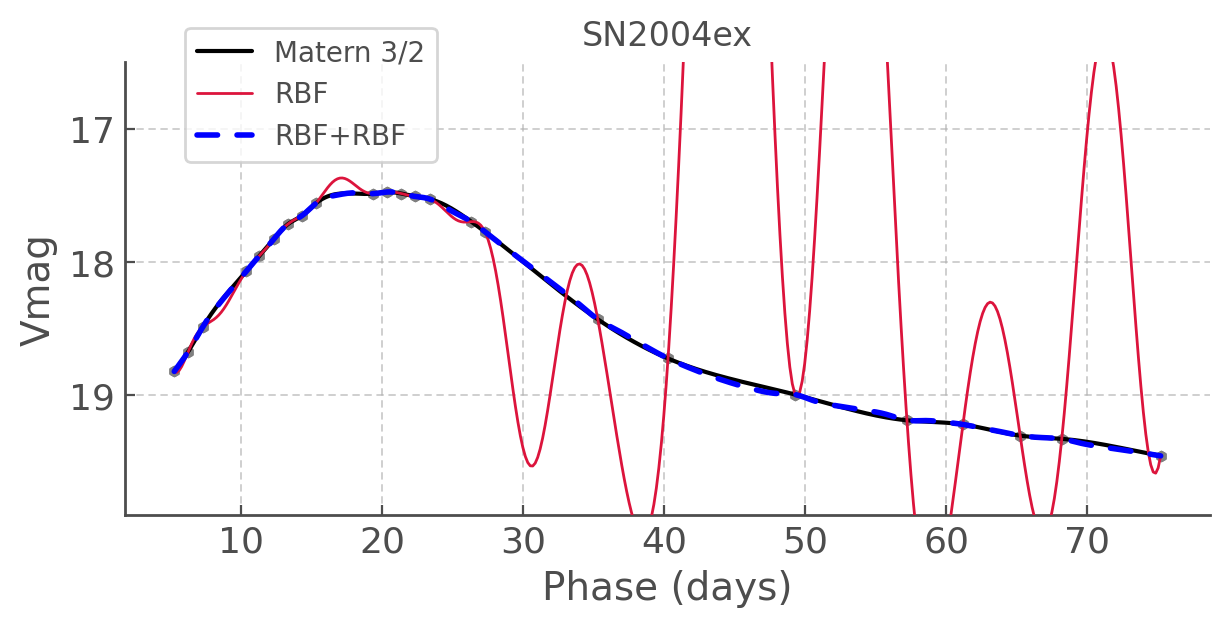}
    \includegraphics[width=7.5cm]{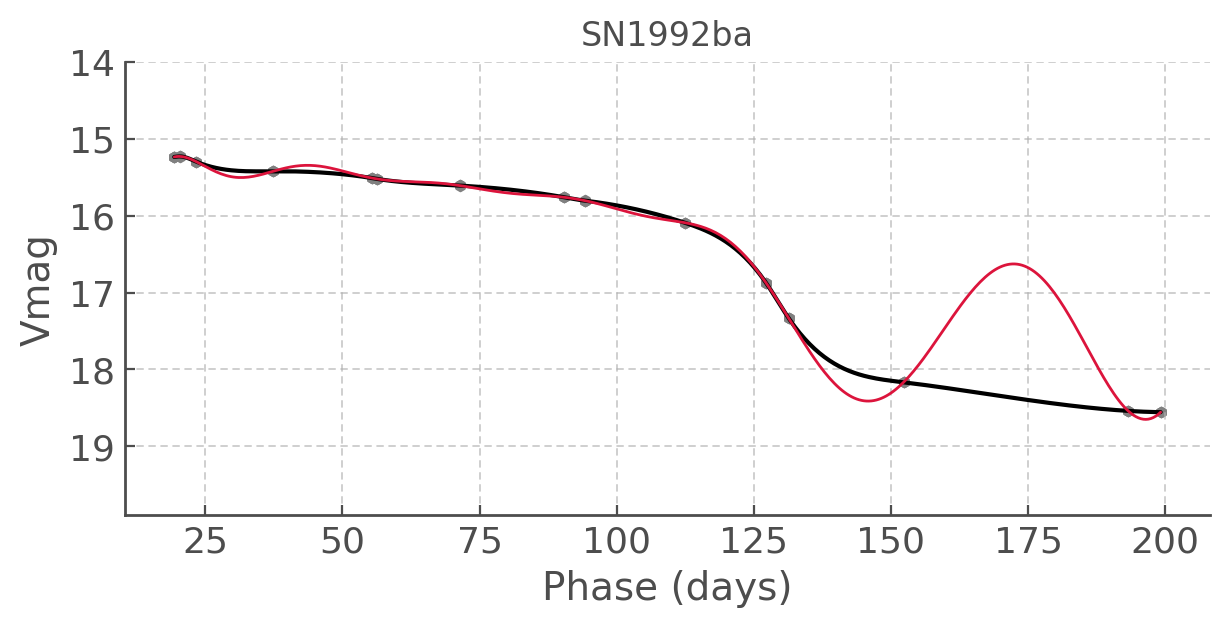}
    \includegraphics[width=7.5cm]{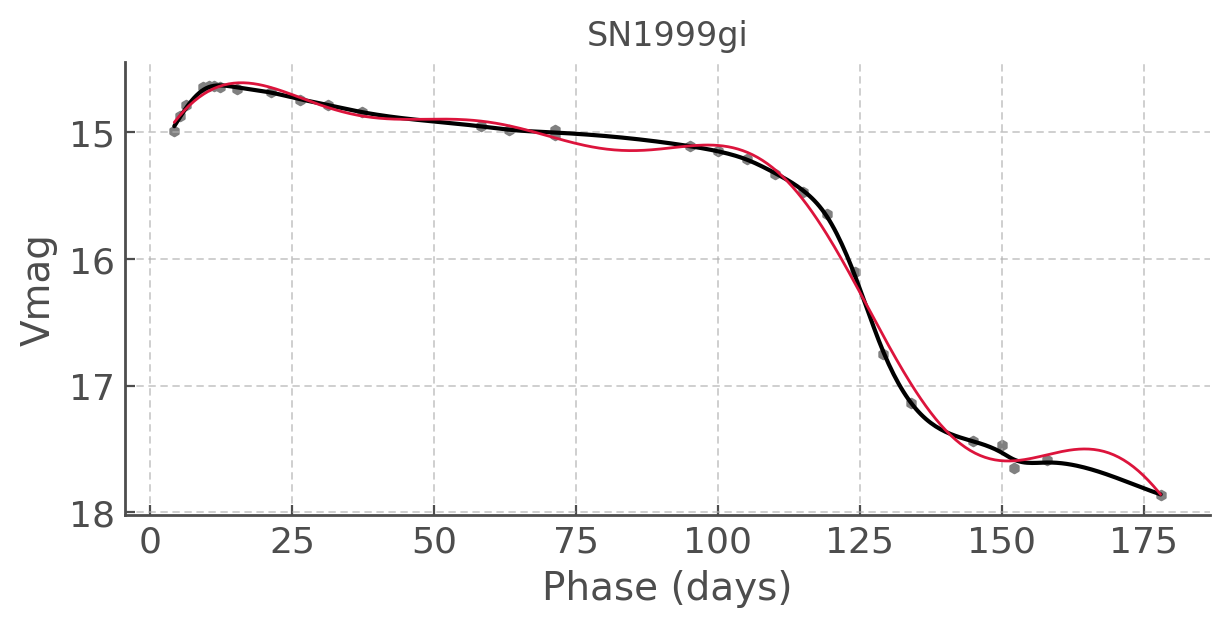}
    \includegraphics[width=7.5cm]{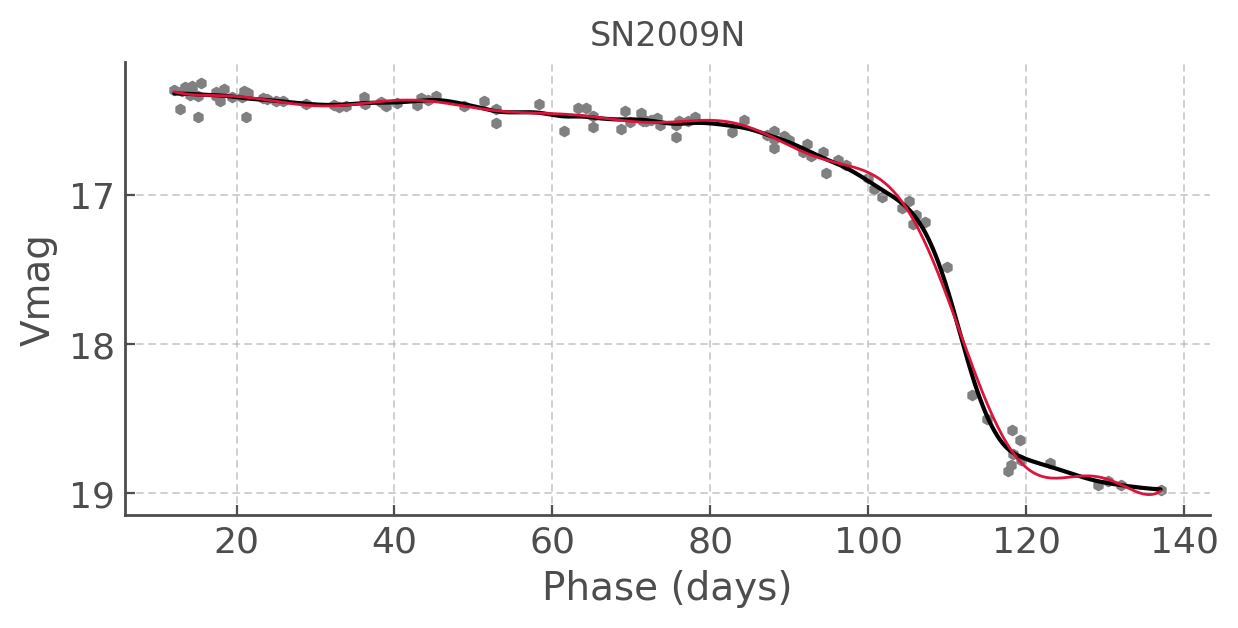}
    \includegraphics[width=7.5cm]{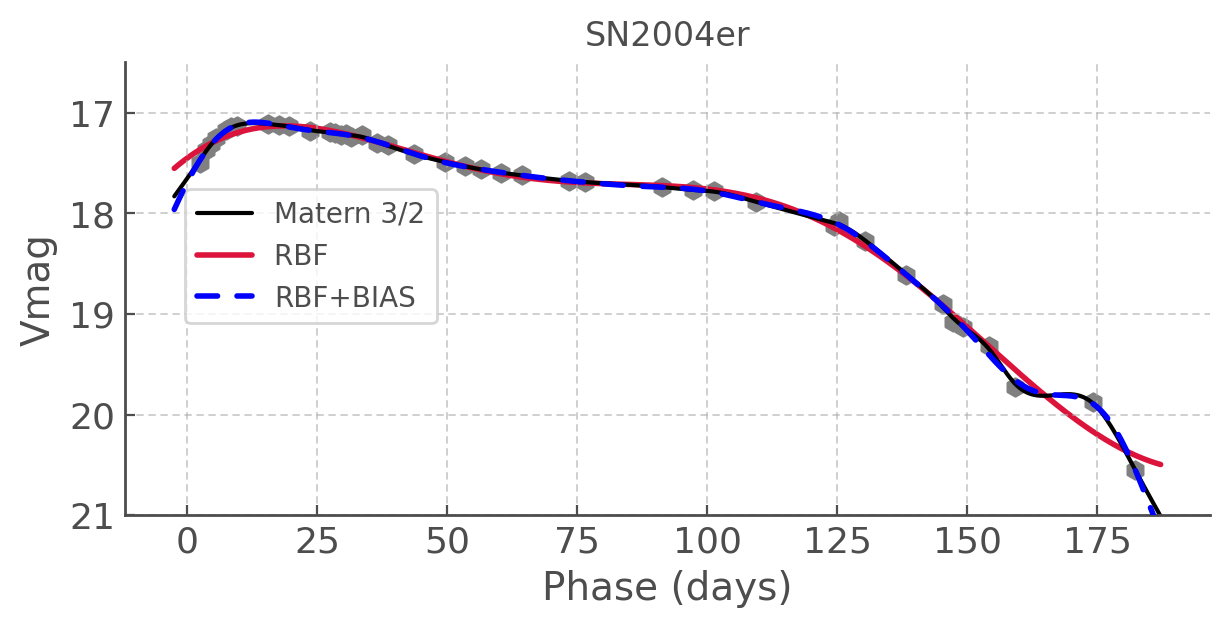}
    \caption{Examples of light curves where the Mat{\`e}rn kernel performs better than the RBF (squared exponential) kernel.}
    \label{fig:gpy_kernel_comp}
\end{figure}
\subsubsection{Mat{\`e}rn 3/2}
\label{sec:matern}
In Figure \ref{fig:gpy_kernel_comp} we show 6 light curves which were not well fitted by a classic RBF kernel and for which the Mat{\`e}rn 3/2 kernel was more appropriate. 
Although in the case of the \cite{boone2019} study they found that the RBF kernel performed practically as well as the  Mat{\`e}rn kernel, we find that for quite a few objects, particularly type IIP SNe, the latter is better suited. 
This is not unexpected, as type IIP are characterised by very slow evolving light curves with a significant drop in magnitude as the plateau phase ends. 
We can see in cases such as SN~1999gi how the RBF kernel is too shallow at the drop of the plateau, an in all cases how much the mean function fluctuates in magnitude when it is physically (and observationaly) expected to be flat. 
The Mat{\`e}rn kernel still displays some unphysical wobble but to a much smaller extent. 

Once again, not all cases are clear cut and the goodness-of-fit question we posed with SN~2011dh is relevant with SN~2004er: the Mat{\`e}rn 3/2 kernel is better at fitting the early rise and peak of the light curve but subsequently over fits the late time light curve. 
The RBF kernel is less prone to over fitting since it does not allow for sharp enough variations in the mean function, but subsequently it cannot fit the rapid rise to peak and in the case of this study where t$_{\rm rise}$ is a parameter of interest, it would be profoundly unsuitable. 

\subsubsection{RBF+RBF}
\label{sec:rbf+rbf}
Another case where the typical RBF kernel failed was SN~2004ex as the GPy converged to a short length scale of 2.9 days leading to large oscillations in the late time light curve and over fitting at early days. 
We suggest that the sparser sampling is to blame in this case, which gives us a window into the difficulties that can be encountered on larger samples of SN light curves with varying quality.

We find that the Mat{\`e}rn 3/2 kernel provided a much more suitable solution, and a similar fit was obtained when combining two RBF kernels that were optimised by the solver to length scales of 27 days and 0.9 days respectively. 
This solution has an interesting physical interpretation: the 27 days length scale kernel relates to the evolution of the light curve from the SN whereas the short length scale kernel accounts for noisy deviations in the individual observations (which are separated by a day or more in these data). 

This is a very interesting prospect for fitting noisier and more sporadic light curves - especially if observing conditions vary from night to night, the level of noise in each data point could vary. 
An additional RBF kernel would be much better suited to capture this point-by-point variability than a white noise kernel. 

\subsection{Full sample fits and length scale considerations}
\label{sec:length_scale}

\begin{figure*}
	\includegraphics[width=18cm]{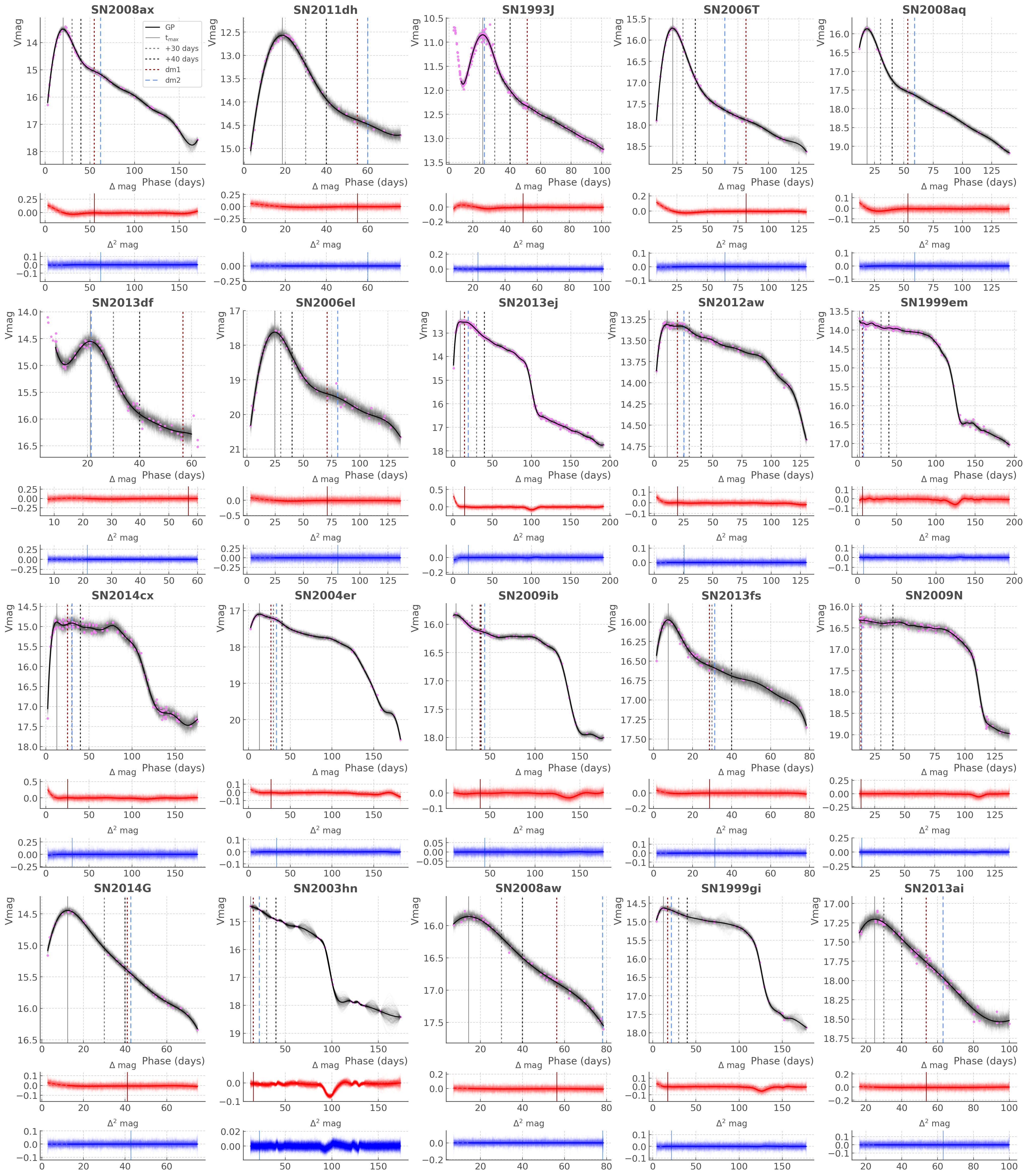}
    \caption{Best GPy regressions of our SN sample -- the kernels used are summaries in Table \ref{tab:gp}. The data is shown in magenta (with error bars which are sometimes smaller than the data points), the mean function of the fit $\mu(\textbf{t})$ is show in black and 100 samples of P(\textbf{m}|\textbf{t}) (Eq. \ref{eq:GP1}) are shown in grey. We also show the first (red) and second (blue) derivatives of $\mu(\textbf{t})$ and of the 100 samples of P(\textbf{m}|\textbf{t}).}
    \label{fig:fullGPy}
\end{figure*}

We now look at the results for our full sample; we will discuss the parameters of interest ($t_{\rm rise}$, $\Delta m_{\rm 40-30}$,  dm1,  dm2), their uncertainties and astrophysical implications in the following sections.
In Table \ref{tab:gp} we record which kernels were used for each SN and in Figure \ref{fig:fullGPy} we show the fits and resulting first and second derivatives. 
Note that for the type IIbs SN~2013df and SN~1993J the adiabatic cooling part of the light curve is fit well by our models but not showed in full here as it is the radioactive decay peak that is used to calculate t$_{\rm rise}$.

\begin{table}
	\centering
	\caption{Kernels used to obtain best fit to each SNe. $\lambda$ is the length scale of convergence for the squared-exponential (RBF) kernels -- see Section \ref{sec:length_scale} for a discussion of the length scales.}
	\label{tab:gp}
	\begin{tabular}{p{1.4cm}p{2.5cm}p{2.5cm}}
		\hline
		Name & GPy kernels & {\tt sklearn} kernels \\
		\hline
		\textbf{Type IIb} \\
		SN 2008ax & RBF ($\lambda=32.4$) & RQ+White \\
		SN 2011dh & RBF ($\lambda=35.5$)  & \\
		SN 1993J & RBF ($\lambda=11.3$) &  RQ+White \\
		SN 2006T & RBF ($\lambda=34.2$) & RQ+RQ  \\
		SN 2004ex & RBF ($\lambda=27.9$) +  RBF($\lambda=0.87$) & RQ+RQ \\ 
		SN 2008aq & RBF ($\lambda=32.6$) & RBF \\
		SN 2013df & RBF ($\lambda=17.7$) & RBF \\
		SN 2006el & RBF  ($\lambda=43.0$) & --\\
		 --  \\
		\textbf{Type II}  \\
        SN 2013ej & RBF ($\lambda=13.4$) & RBF \\
        SN 2012aw & RBF ($\lambda=37.2$) & RBF \\
        SN 1999em & Mat{\`e}rn--3/2 & RBF \\
        SN 2014cx &  RBF ($\lambda=21.6$) & RQ + White\\
        SN 2004er & RBF ($\lambda=16.2$) + Bias & RBF \\
        SN 2009ib & RBF  ($\lambda=25.1$) &  RBF\\
        SN 2013fs  & RBF ($\lambda=23.1$) & RBF \\
        SN 2009N  & Mat{\`e}rn--3/2 & RQ+ White \\
        SN 2014G  & RBF ($\lambda=32.7$) & RBF \\
        SN 2003hn &  Mat{\`e}rn--3/2 & RQ + White \\
        SN 2008aw & RBF ($\lambda=56.7$) & RBF \\
        SN 1999gi & Mat{\`e}rn--3/2 & RBF \\
        SN 1992ba & Mat{\`e}rn--3/2  & RBF\\
        SN 2013ai & RBF ($\lambda=42.9$) & RBF \\
		\hline
	\end{tabular}
\end{table}

When looking at Figure \ref{fig:fullGPy} it is clear that GP models have a tendency to over-fit the SN light curves.
In the case of type IIb SNe that is mostly the case at later times on the radioactive decay tail; in type II SNe it occurs both on the plateau and the tail. This is particularly true for some of the more difficult to fit data sets such as SN~2014cx, but even in the case of SN 2008ax which is generally very successful before 150 days, a slight "wobble" can be seen in the data. 
This type of oscillation can be seen in figure 1 of \citetalias{pessi2019} as well as in their figure 2 (although the over plotting makes is harder to distinguish), and it is not user error.

This is a \textit{direct and inevitable} consequence of the commonly used kernels in GP regression: they are stationary kernels and as a result intrinsically fit for a function with \textit{unchanging} length scale over the course of the event.
We know that not to be the case with SNe.
In light curves with a clear radioactive peak and decay tail, the evolution of the peak powered by Nickel-56 ($\tau_{1/2}=6.1$ days) will be faster than that of the Cobalt-56 ($\tau_{1/2}=77$ days) powered tail. 
In type IIP SNe the plateau phase can last for a few months, but the hydrogen recombination and fall off the plateau lasts a few weeks at most. 
By adding kernels together we can account for the behaviour of complex functions that evolve on several timescales \textit{simultaneously} (e.g. the mixture of two RBF kernels for SN~2004ex) but our GP models cannot appropriately reproduce a function that has timescales which change over time.

In Table \ref{tab:gp} we report the convergence length scales from the GPy fits. 
We can see that they are typically between 15 and 40 days, which is the time window in the early light curve or in the drop off the plateau where the most change will occur.
These phases show the most amplitude variation in the light curve and are the most well sampled (at least in this data set) -- consequently during the optimization phase and the maximisation of the log likelihood, the model will be penalised the most if it does not reproduce these portions of the light curve well.
That is why they will be the dominant driver of the length scale of convergence. 

Now in general our GPs and the basic kernels perform sufficiently well when the data is well sampled (which is the case for most of our SNe by design) as it anchors the fits: a good example of this is SN~2013ej, which has a convergence length-scale of 13.4 days which is much shorter than that time scales on which most of its light curve varies in smoothness, but the dense sampling anchors the models and the small length scale only results in minor oscillations that are withing uncertainties.
But cases such as SN~2003hn and SN~1999gi show what the impact of too small a length scale can have on re-sampling too sparse a data-set: a `bloated' variance between the data points, which does not reflect the likely variations in the light curve at these epochs.

In \citetalias{pessi2019} the average time between observations is used as the prior length scale: as can be seen from Table \ref{tab:gp}, this does not reflect the typical length scales of our SN light curves.
The issue with using the average time between observations as the length scale is not simply that it is not related to the evolution of our SNe, it is also because in practice light curve sampling is most dense at early days.
Therefore the average time between samples will typically be lower than the time separation of the data in the tail (which is already prone to over-fitting even with dense sampling), and if the length scale is smaller than the separation between the data points very large oscillations can occur in the GP solution as we saw above in the cases of SN~2003hn and SN~1999gi. 
Therefore this length scale prescription has the potential to worsen the over-fitting of the late time light curve depending on the implementation of GP used (see Section \ref{sec:sklearn}).
We surmise that \citetalias{pessi2019} encountered no major issues because they used GPy which converges to reasonable values most of the time even with unphysical initial guesses, as we tested above.

Strictly speaking, the fitting of SN light curves with GP should require the use a \textit{non-stationary} kernels which are capable of fitting functions with variable smoothness (e.g. see Figure 1 of \citealt{paciorek2003}).
Such and implementation is beyond the needs or scope of this paper, but we emphasise that this detail in the kernel choice could be highly relevant for teams looking to fit large samples of SN light curves of varied types and they are set on using GP regression.

\subsection{Different implementations, different kernels}
\label{sec:sklearn}

\begin{figure}
	\includegraphics[width=7cm]{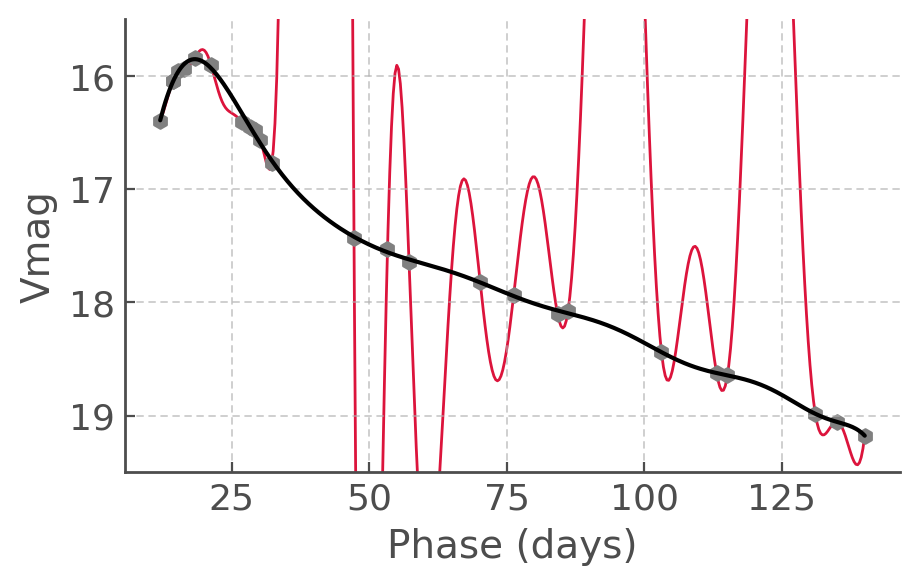}
    \caption{Red curve: {\tt sklearn} GP regression with RBF kernel with length scale bounds (mean time interval between epochs; 100 days). Black curve: {\tt sklearn} GP regression with RBF kernel with length scale bounds (30, 100).}
    \label{fig:08aq}
\end{figure}

A final point of discussion related to Gaussian processes that we want to address is that different implementations can yield different results.
To test this we fitted our whole sample with {\tt sklearn} and recorded which kernels performed best in Table \ref{tab:gp} - we do not show the final mean fits themselves as for the most part they are similar (although not the same) as those presented in Figure \ref{fig:fullGPy}.
However as an example we show GP regression performed on the light curve of SN~2008aq in two different configurations (see Figure \ref{fig:08aq}). 
The first (red) fit was performed using a length scale bounds starting at the mean time interval between epochs (a few days) as we did throughout our GPy fits, whilst the latter had a lower bound of 30 days, which is more typical as seen in Table \ref{tab:gp}.
In the GPy implementation we have seen that the start length scale had little influence on the final value of convergence, but here the lower length scale bound resulted in a model that severely over-fit our data -- this behaviour was noted across our sample. 

Generally we find that the RBF kernels in GPy and {\tt sklearn} behave slightly differently and in the latter fits are sometimes more successful when performed with a Rational Quadratic (RQ) kernel (essentially a sum of RBF kernels). 
Additionally the Mat{\`e}rn--3/2 kernel in {\tt sklearn} never performed better than the RBF or the RQ kernels.
It is unclear why that is the case and in terms of analysis and results it is not particularly important: so long as the fits are close enough to the data (within uncertainties) the re-sampling will be successful. 
This consideration does have an impact on how we share our findings and what `reproducibility' means for this type of work when using so-called black boxes: it is important to be specific about the packages employed and, if possible, to release publicly the codes and scripts used for analysis.


\section{Ligth curve Parameters}
\label{sec:results}
We now turn our attention to the light curve morphology parameters $t_{\rm rise}$, $\Delta m_{\rm 40-30}$, dm1 and dm2.
In this section we review the uncertainties associated with these parameters in the context of GP fits, and discuss which ones can reliably be used for astrophysical interpretation . 

\subsection{dm1,  dm2 and uncertainties}
\label{sec:uncertainties}
The issues of over fitting and the presence of oscillations in the model that are not present in the light curve which we discussed above are causes for concern for the values and uncertainties on the first and second derivatives. 
In Figure \ref{fig:fullGPy} we show the first and second derivatives of 100 re-samplings of the models for each SN  over 400 epochs - we also show the position of the maximima of the first and second derivatives of the mean function $\mu(\textbf{t})$.
It is this mean function that \citetalias{pessi2019} use for their calculation and that they plot in their figure 3, but it is important to re-sample the models using the posterior co-variances calculated by the optimizer. 
As we can see the level of noise is such that the position of the first maxima in the first and second derivatives of the mean functions are not at all obvious.

\begin{figure}
	\includegraphics[width=8.5cm]{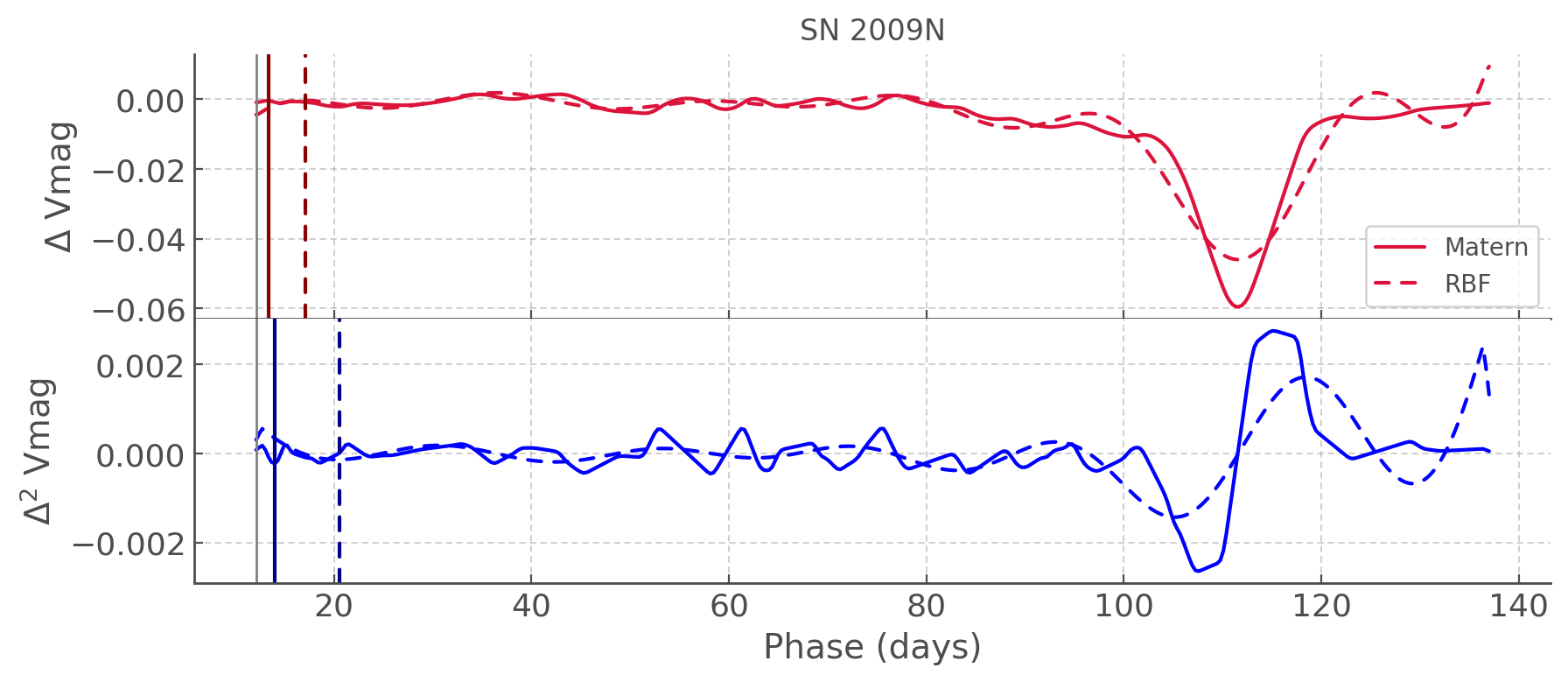}
	\includegraphics[width=8.5cm]{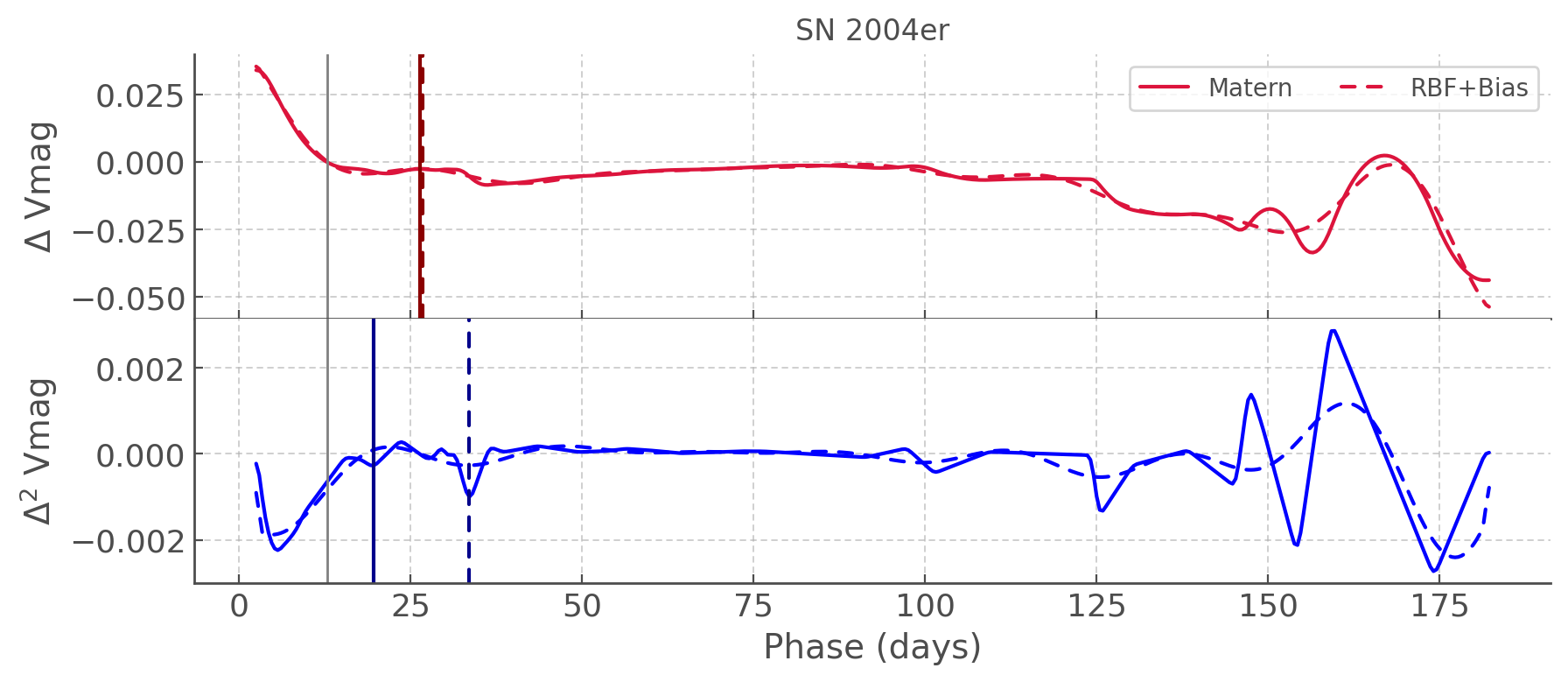}
	\includegraphics[width=8.5cm]{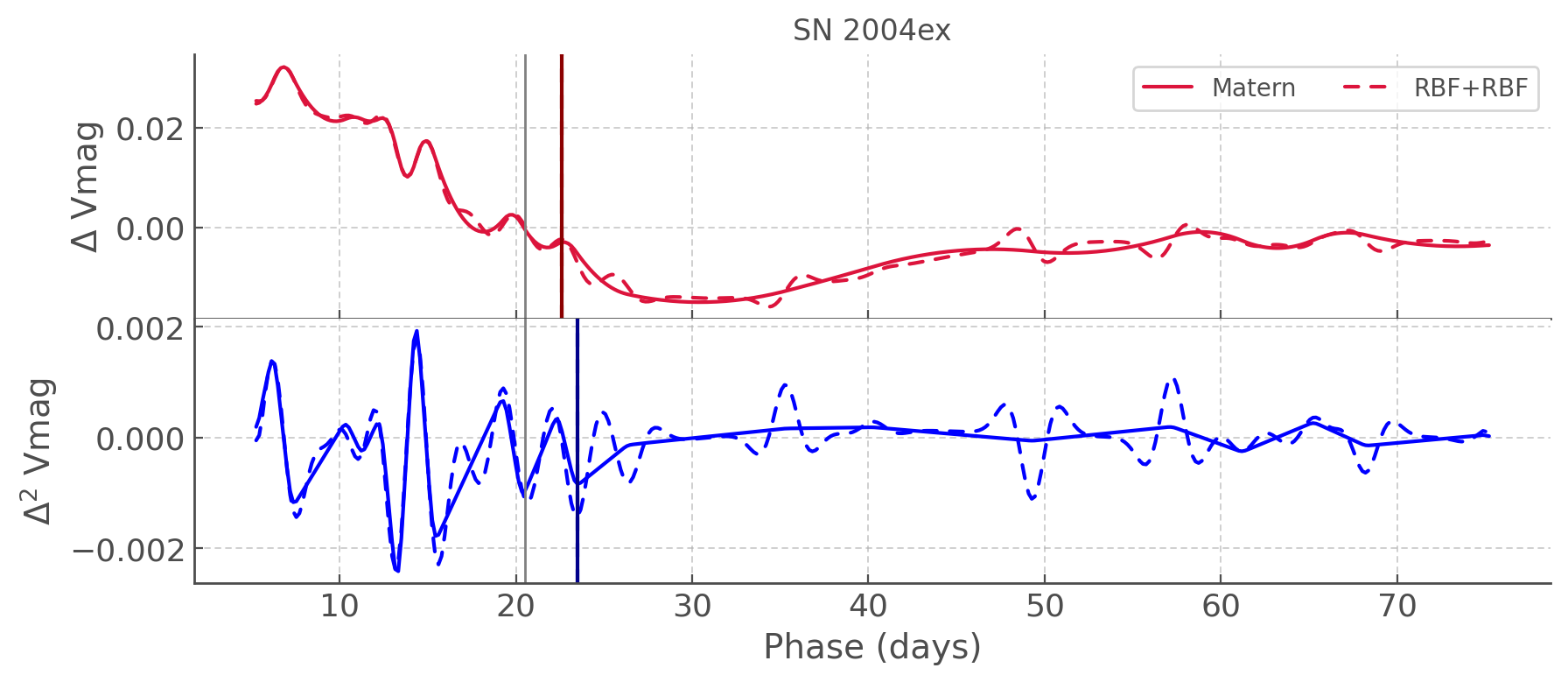}
    \caption{Firs (red) and second (blue) derivatives of the mean functions $\mu(\textbf{t})$ of the GP regressions performed with two sets of kernels (see legends) for SN 2009, SN~2004er and SN~2004ex. The parameters dm1 and dm2 are shown as vertical lines of the style corresponding to the kernels used for the interpolation, and the thin grey lines show the time at maximum light. }
    \label{fig:dm1dm2_diff_kernels}
\end{figure}

Additionally we want to highlight how the derivatives can vary with kernel choice  and how the impact can be unpredictible even when only considering mean values of the parameters.
In Figure \ref{fig:gpy_kernel_comp} we can see the fits of SN~2004er, SN~2004ex and SN2009N resulting form different kernels, in particular a Mat{\`e}rn--3/2 kernel and an RBF(+bias) kernel resulting in fits that are very similar.
But in Figure \ref{fig:dm1dm2_diff_kernels} we show the first and second derivatives of the mean functions for these different kernels and they are strikingly different. 
These differences can have no impact on the parameters as in SN~2004ex where the amplitude of the derivatives changes but their maxima and minima are very similar particularly at early days when the parameters are calculated.
However, this is not always the case. For SN 2004er, the first derivatives are very similar and their resulting dm1 parameters indistinguishable, but the dm2 parameter varies drastically. Finally in the case of SN 2009N, both the dm1 and dm2 parameter are affected by the choice of kernel, despite both fits lying on top of each other until 80 days post-explosion. 
Overall we conclude that using the derivatives of SN light curve fits done with GP is not a reliable way to look for morphological differences related to their progenitors as the results are very unstable to the choice of kernel.

\subsection{$t_{\rm rise}$, $\Delta m_{\rm 40-30}$ and uncertainties}

Although dm1 and dm2 suffer too much intrinsic uncertainty, the rise time $t_{\rm rise}$ and magnitude difference between phase 40 and 30 $\Delta m_{\rm 40-30}$ are well determined given sufficient sampling.
 
 \begin{figure}
	\includegraphics[width=8.5cm]{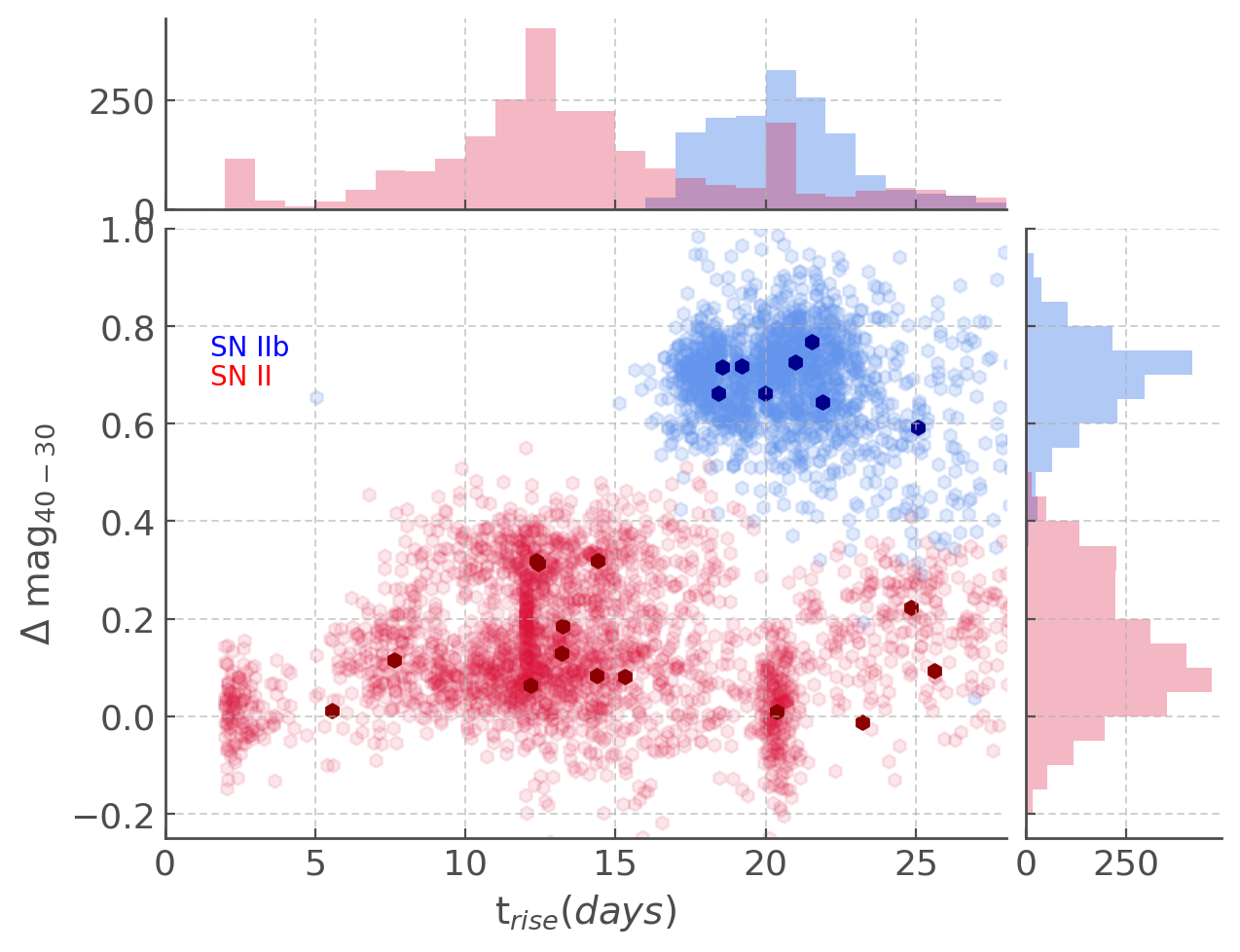}
    \caption{$t_{\rm rise} - \Delta m_{\rm 40-30}$ parameter space showing the values of the parameters for 100 re-sampling of each SN model. The type II SN are show in red and the type IIb in blue. The darker points show the values of $t_{\rm rise}$ and $\Delta m_{\rm 40-30}$ for the mean functions of each SN model.}
    \label{fig:trise_dmag}
\end{figure}

\begin{figure}
\centering
	\includegraphics[width=8.5cm]{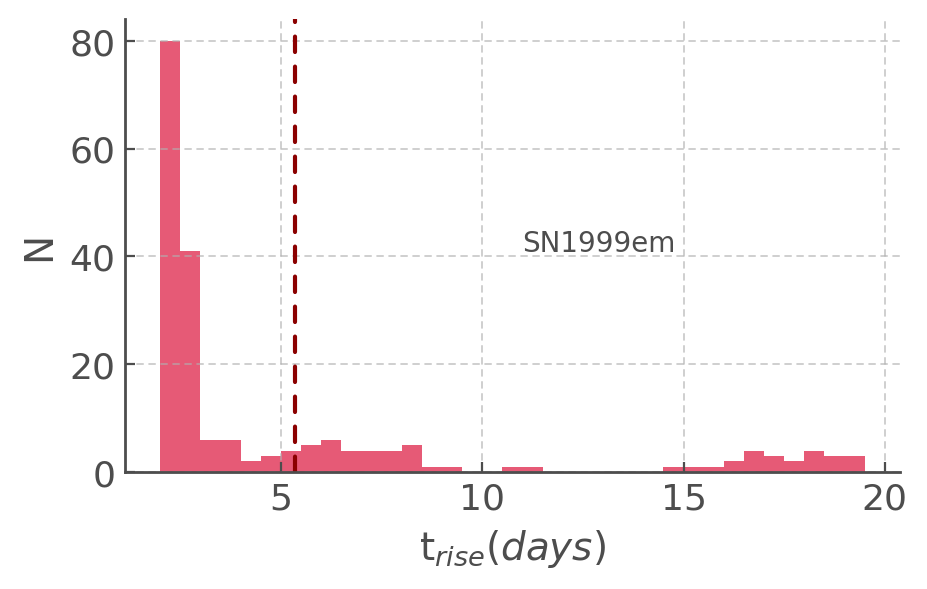}
    \caption{Distribution of the value of t$_{\rm rise}$ in days for 100 samples of the model of SN1999em. The mean is shown as the dashed line.}
    \label{fig:99em_trise}
\end{figure}

In Figure \ref{fig:trise_dmag} we show the $t_{\rm rise} - \Delta m_{\rm 40-30}$ parameter space. We do not use error-bars to represent our uncertainties as they are for the most part non-Gaussian. Instead we re-sample the GP models 100 times per SN and record the two parameters. 
We also plot the values of  $t_{\rm rise}$ and $\Delta m_{\rm 40-30}$ calculated for the mean functions of each SN. 
It is worth noting that the mean will not necessarily be found were the density of points is highest in many cases -- an example is SN~1999em where the mean is very offset from the majority of the data due to the shape of its distribution (see Figure \ref{fig:99em_trise}). 
It is because of this behaviour that we chose re-sampling over the plotting of the mean and standard deviation.

Overall the $\Delta m_{\rm 40-30}$ distributions for type II and IIb SNe are distinct but adjacent.
The small sample of SNe here, and the use of only one light curve re-sampling method does not allow us to draw definitve conclusions on the shape of these distributions, but it suggests the divide may not be as marked as presented in \citetalias{pessi2019}.
Calculating these parameters for a much more exhaustive sample will be necessary and we suggest that data from the ATLAS transient survey \citep{smith2020} would be ideal for this purpose:  the <1 day cadence in the light curves would provide dense enough sampling that complex fitting and re-sampling would not be required to obtain values of $t_{\rm rise}$ and $\Delta m_{\rm 40-30}$ with small uncertainties. 
The consistent cadence would make these data easy to fit with GP models as done here, although alternatives, such as Bazin analytical models \citep{bazin2009, villar2019}, would likely be preferable due to the known over-fitting issue discussed in earlier sections.

Finally we note that Figure 6 does not take into account uncertainties on the explosion date, which would affect $t_{\rm rise}$. This was deliberate to focus on the natural spread that arises from GP modeling, and given the present results our conclusions would be unchanged. Indeed, the explosion date uncertainty is at most a few days (see Figures \ref{fig:LC_IIb} and \ref{fig:LC_II}) and the peaks of the $t_{\rm rise}$ distributions for the type II and IIb supernovae are \about 10 days apart, so they would remain clear; additionally those distributions already overlap, and the explosion date uncertainty would only strength this overlap. We do caution that taking into account the explosion date uncertainty in this type of analysis would be non-trivial: the explosion date would need to be sampled, but the appropriate distribution is not clear as the errors on the explosion dates are systematic rather than stochastic. An extended discussion of this challenge is beyond the scope of this publication. 

\section{Discussion}
\label{sec:discussion}
The parameters defined by \citetalias{pessi2019} t$_{\rm rise}$, $\Delta m_{\rm 40-30}$, dm1 and dm2 each are intended to capture a characteristic of light curve morphology which can lead to physical interpretation. 
dm1 and dm2, the first maxima post peak of the first and second derivatives of the light curve, should indicate the maximum rate at which the SN fades and how much the curvature of the light curve changes where it changes the most.
As we have seen in the previous section, the over-fitting from GP and the uncertainty in the re-sampled gradients make these parameters impractical for astrophysical interpretation in this particular context, but they may be more successful when calculated with an alternative fitting method.

t$_{\rm rise}$ and $\Delta m_{\rm 40-30}$, the time from explosion to maximum light and the magnitude difference between phase 40 and 30 (post-explosion) respectively, have much more stable values and we do see in our results a clear clustering of the type II and IIb.
This separation between type II and IIb SNe in parameter space was more severe in \citetalias{pessi2019} and interpreted as potentially caused by separate progenitor channels as the light curve decline in particular is correlated with how much hydrogen remains in the envelope at the time of explosion: a lesser degree of envelope stripping leaves behind more hydrogen which can contribute more light on recombination providing an excess compared to a pure radioactive decay tail.

A distinction between the two classes had been suggested by \cite{arcavi2012} who reported that rapidly declining light curves in their sample solely belonged to type IIb SNe.
However the sample was small (20 SNe) and transitional events showing characteristics of both type II and IIb SNe have now been reported (SN~2013ai; \citealt{davis2021}).
Additionally, \cite{anderson2019} performed a meta-analysis of CCSNe, their Nickel-56 yields, and implications for their progenitors; they reviewed existing evidence from the literature which point to an interesting dichotomy:  Based on direct imaging of the progenitors, type II and stripped envelope SNe do not seem particularly difference, whereas from environment studies seem there may be a decreasing age sequence from type II to more stripped sub-types. 

In this context, should we expect further studies of t$_{\rm rise}$ and $\Delta m_{\rm 40-30}$ to show type II and IIb as clear spectrum or should we expect two distinct clusters? 
Here we flip the problem on its head and use results from theoretical models of binary stellar evolution to qualitatively assess what morphologies we expect to see.
To do so we use the BPASS fiducial models which have been extensively tested against observations and are capable of reproducing a large array of observable phenomena particularly in massive stars (e.g. \citealt{eldridge2017,stanway2018, rosdahl2018,eldridge2019,stanway2020a,stanway2020b,massey2021,ghodla2022,briel2022,stevance2022}).
We use {\tt hoki} \citep{stevance2020} to search for type II and IIb progenitors using similar criteria for the hydrogen-rich/hydrogen-poor boundary as in \cite{stevance2021}:  H-rich progenitors are taken as containing more than 0.001 \msol of hydrogen at time of death with a hydrogen mass fraction greater than 10 percent \citep{dessart12}.
Also we only consider models which will explode as SNe by searching for models with a CO core mass at death be greater than 1.38 \msol and that the remnant be less than 3\msol\footnote{baryonic mass}.
Although the type IIb classification itself is unlikely to have a sharp threshold in envelope hydrogen mass, we use the commonly used 0.5\msol values \citep{smith2011} as a boundary between type II and type IIb.

\begin{figure}
	\includegraphics[width=8cm]{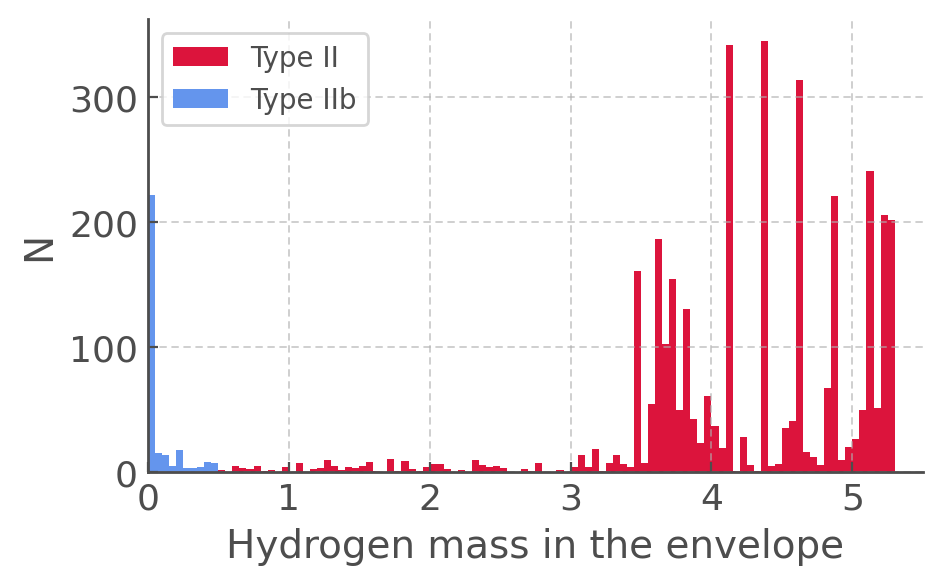}
    \caption{Hydrogen mass remaining in the envelope at the time of explosion for the type IIB and II progenitors in BPASS at Z=0.020.}
    \label{fig:IIb}
\end{figure}

In Figure \ref{fig:IIb} we show the number of type II and IIb SN in a 10$^6$\msol stellar population born with Z=0.020 (solar metallicity) as function of remaining hydrogen mass. 
We see that their is no gap between the type II and IIb SNe, but there is a large number of systems with very low hydrogen envelope masses ($<$0.05\msol) whereas the majority of type II explosions have $>3.5$\msol of hydrogen left in their envelope as they explode. 
This would explain why with low number of observations the two classes could look quite distinct and with growing sample intermediary events are beginning to emerge. 

\begin{figure}
	\includegraphics[width=8cm]{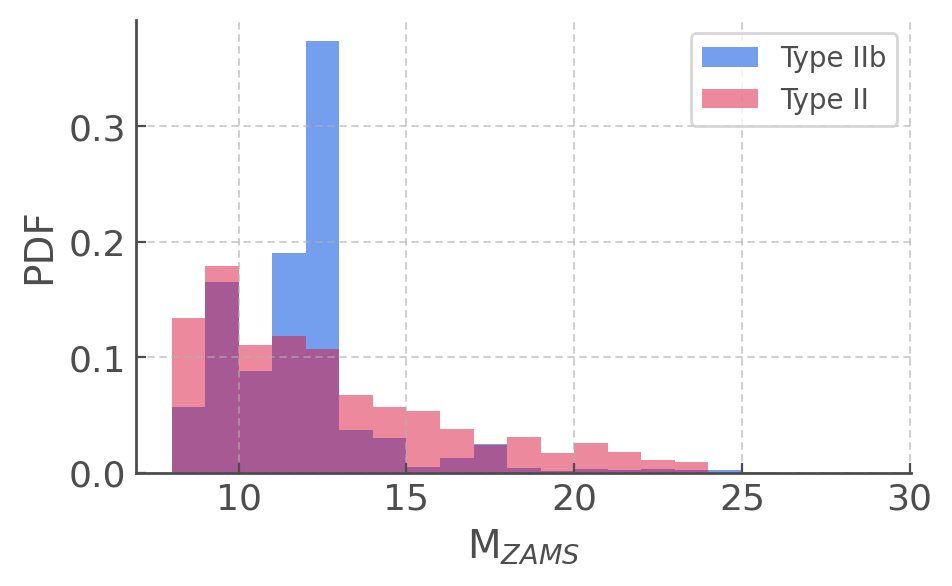}
    \caption{Zero Age Main Sequence mass of the type II and IIb SN progenitors in BPASS at Z=0.020}
    \label{fig:IIb_ZAMS}
\end{figure}

Although it is not the object of this paper to draw any firm conclusions on progenitor differences we can begin to explore our model population by looking at the distribution of ZAMS masses (see Figure \ref{fig:IIb_ZAMS}). 
We can see that type II ZAMS masses steadily decline as we move to high masses as we expect from the point of view of the initial mass function,
 the type IIb on the other hand show a density peak around 12 solar masses.
This behaviour could explain the dichotomy in the observations: host environments could tend to indicate that type IIb SNe come from more younger stars because their ZAMS mass distribution is skewed to higher values (and therefore lower ages), but since the type II ZAMS mass distribution spans that of the type IIb direct detection of progenitors would be virtually identical because of low-number statistics.

On the whole the answer to the question "are type II and IIb progenitors different?" is likely a mixture of `yes' and `no', if only because the classification is observational and unlikely to represent a systematic divide in the stellar evolution.
Nevertheless we can look to our fiducial BPASS stellar population to find differences of note in the genealogies of the type IIb and II SNe in our simulations. 
We find here that 100 percent of the type IIb were at one point in a binary system, whereas the type IIs come from single stars in \about 30 percent of cases. 
Interestingly we also find that nearly 50 percent of the type IIb progenitors are merger products, whereas this value is only \about 16 percent for type IIs.
The importance of binary interactions in order to strip the progenitors of type IIbs has long been established (e.g. \citealt{podsiadlowski1992,podsiadlowski1993, nomoto1993, young2006, stancliffe2009, maund2009, clayes2011,yoon2017}),
but further analysis will be required to understand how these differences in the proto-typical evolutionary channels relate to the observable morphologies of type II(b) light curves, and it may also be the case that the genealogies of the most heavily stripped type IIbs differ from that of events which are closer to the type II boundary: we may see a spectrum of differences in the stellar evolution as well as in the morphologies.

\section{Conclusions}
\label{sec:conclusion}
In this paper we used 22 of the best sampled type II(b) SNe to carefully investigate the suitability of Gaussian Process regression in the context of SN light-curves generally and in re-sampling the data to determine the morphological parameters t$_{\rm rise}$, $\Delta {\rm mag}_{\rm 40-30}$ defined by \cite{pessi2019}.

We find that due to the the way GPs are trained the light curve peak or the drop of the plateau will be dominant factors in the length scale optimisation.
Since these parts of the light curve evolve on shorter times scales than others (e.g. late tail, plateau), the slower phases will systematically be over fit.
On the whole GPs with traditional kernels (squared exponential \citealt{kim2013,pessi2019}; Mat{\`e}rn--3/2  \citealt{boone2019}) are technically speaking not suitable to reproduce SN light curves due to the varying timescale of the phenomenon. 
An avenue would be to implement GP regression with non-stationary kernels \citep{paciorek2003} that can reproduce data with varying smoothness.

In general if the main goal of GP regression is to fit solely the early time light curve rather than the time series in full, the issue of varying time scales remains minor.
Additionally if implementing non-stationary kernels is not feasible we would none the less recommend testing for the best kernel combination, which will be dependent on the sampling and shape of the light curves in a given data set. 
For example we found the Mat{\`e}rn--3/2 kernel to perform better than the typical squared-exponential (RBF) kernel on several occasions.
Quite notably we discovered that the combination of two RBF kernels could be powerful in capturing the SN light curve behaviour as well as point-to-point deviations (see Section \ref{sec:rbf+rbf}) -- since the noise in the data is likely to be different from white noise in most cases this kernel combination could be useful.
Finally it is worth pointing out that the best set-up for GP regression is to a large degree dependent on the implementation (see Section \ref{sec:sklearn}).

On the whole, good fits can be found with GP regression and re-sampling given that the data are well sampled to anchor the fit. 
In the context of this study this means that for parameters like t$_{\rm rise}$ and $\Delta {\rm mag}_{\rm 40-30}$ the slight over fitting does not increase the uncertainties in most cases (in the case of type IIPs whose early light curve would be best fit by a straight line however the t$_{\rm rise}$ values from GPs are mostly determined by the over-fitting). 
The main issue is in the calculation of the gradients from the models. 
We find that the large uncertainties and the tendency for the gradients to be sensitive to the kernel even when the mean fits are indistinguishable makes these parameters unreliable for astrophysical interpretation in the present case.
They might however be worth considering if analytical fits not prone to "wobbliness" were performed instead.

We do find clear clustering of the type II and IIb SNe in t$_{\rm rise}$ -- $\Delta m_{\rm 40-30}$ space, although the division between the two classes is not as clear as in \citetalias{pessi2019}, possibly because of the use of re-sampling rather than error bars in our visualisation. 
Further study of t$_{\rm rise}$ and $\Delta m_{\rm 40-30}$ using a larger sample of events will be required to draw firm conclusions on this parameter space. 
We propose that the ATLAS transient survey \citep{smith2011} will provide the ideal data set for this purpose given the dense sampling (<1--2 days between observations over the whole light curve). 
We will also test the scalability of fitting with analytical models (e.g. \citealt{villar2019}) to derive the light curve morphologies.

Given the type II and IIb progenitors predicted by the BPASS fiducial models we expect the t$_{\rm rise}$ -- $\Delta m_{\rm 40-30}$ parameter space to show a bimodal behaviour with a number of transitional objects bridging the gap between the clear type IIb SNe with hydrogen envelope masses $<$0.1\msol and clear type II(P) progenitors with hydrogen envelope masses $>$3.5\msol.

\section*{Acknowledgements}
The authors thank the referee for their careful review of our manuscript. HFS is very thankful to P.J. Pessi for forwarding data and code which allowed a smoother comparison to their original study. 
HFS acknowledges the Marsden Fund Council managed through the Royal Society of New Zealand Te Apārangi as well as the University of Auckland Summer Scholarship N$^{\rm o}$ SCI008 which funded the work of AL. 
HFS would like to thank S. Littlefair for very helpful and animated discussions about Gaussian Processes and J.J.Eldridge for her support and mentoring.

\section*{Data Availability}
The supernovae light curves are presently available at the Open Supernova Catalogue. 
The data and code used to perform the analysis presented in this publication can be found on GitHub in the following repository: \url{https://github.com/HeloiseS/CODE_AND_DATA_MN_22_2485_MJ}



\bibliographystyle{mnras}
\bibliography{bib} 

\begin{thebibliography}{}
\makeatletter
\relax
\def\mn@urlcharsother{\let\do\@makeother \do\$\do\&\do\#\do\^\do\_\do\%\do\~}
\def\mn@doi{\begingroup\mn@urlcharsother \@ifnextchar [ {\mn@doi@}
  {\mn@doi@[]}}
\def\mn@doi@[#1]#2{\def\@tempa{#1}\ifx\@tempa\@empty \href
  {http://dx.doi.org/#2} {doi:#2}\else \href {http://dx.doi.org/#2} {#1}\fi
  \endgroup}
\def\mn@eprint#1#2{\mn@eprint@#1:#2::\@nil}
\def\mn@eprint@arXiv#1{\href {http://arxiv.org/abs/#1} {{\tt arXiv:#1}}}
\def\mn@eprint@dblp#1{\href {http://dblp.uni-trier.de/rec/bibtex/#1.xml}
  {dblp:#1}}
\def\mn@eprint@#1:#2:#3:#4\@nil{\def\@tempa {#1}\def\@tempb {#2}\def\@tempc
  {#3}\ifx \@tempc \@empty \let \@tempc \@tempb \let \@tempb \@tempa \fi \ifx
  \@tempb \@empty \def\@tempb {arXiv}\fi \@ifundefined
  {mn@eprint@\@tempb}{\@tempb:\@tempc}{\expandafter \expandafter \csname
  mn@eprint@\@tempb\endcsname \expandafter{\@tempc}}}

\bibitem[\protect\citeauthoryear{{Anderson}}{{Anderson}}{2019}]{anderson2019}
{Anderson} J.~P.,  2019, \mn@doi [\aap] {10.1051/0004-6361/201935027}, \href
  {https://ui.adsabs.harvard.edu/abs/2019A&A...628A...7A} {628, A7}

\bibitem[\protect\citeauthoryear{{Anderson} et~al.,}{{Anderson}
  et~al.}{2014}]{anderson2014}
{Anderson} J.~P.,  et~al., 2014, \mn@doi [\apj] {10.1088/0004-637X/786/1/67},
  \href {https://ui.adsabs.harvard.edu/abs/2014ApJ...786...67A} {786, 67}

\bibitem[\protect\citeauthoryear{{Arcavi} et~al.,}{{Arcavi}
  et~al.}{2012}]{arcavi2012}
{Arcavi} I.,  et~al., 2012, \mn@doi [\apjl] {10.1088/2041-8205/756/2/L30},
  \href {https://ui.adsabs.harvard.edu/abs/2012ApJ...756L..30A} {756, L30}

\bibitem[\protect\citeauthoryear{{Barbon}, {Benetti}, {Cappellaro}, {Patat},
  {Turatto}  \& {Iijima}}{{Barbon} et~al.}{1995}]{barbon1995}
{Barbon} R.,  {Benetti} S.,  {Cappellaro} E.,  {Patat} F.,  {Turatto} M.,
  {Iijima} T.,  1995, \aaps, \href
  {https://ui.adsabs.harvard.edu/abs/1995A&AS..110..513B} {110, 513}

\bibitem[\protect\citeauthoryear{{Bazin} et~al.,}{{Bazin}
  et~al.}{2009}]{bazin2009}
{Bazin} G.,  et~al., 2009, \mn@doi [\aap] {10.1051/0004-6361/200911847}, \href
  {https://ui.adsabs.harvard.edu/abs/2009A&A...499..653B} {499, 653}

\bibitem[\protect\citeauthoryear{{Benson} et~al.,}{{Benson}
  et~al.}{1994}]{benson1994}
{Benson} P.~J.,  et~al., 1994, \mn@doi [\aj] {10.1086/116958}, \href
  {https://ui.adsabs.harvard.edu/abs/1994AJ....107.1453B} {107, 1453}

\bibitem[\protect\citeauthoryear{{Bianco} et~al.,}{{Bianco}
  et~al.}{2014}]{bianco2014}
{Bianco} F.~B.,  et~al., 2014, \mn@doi [\apjs] {10.1088/0067-0049/213/2/19},
  \href {https://ui.adsabs.harvard.edu/abs/2014ApJS..213...19B} {213, 19}

\bibitem[\protect\citeauthoryear{{Blondin} \& {Tonry}}{{Blondin} \&
  {Tonry}}{2007}]{blondin2007}
{Blondin} S.,  {Tonry} J.~L.,  2007, \mn@doi [\apj] {10.1086/520494}, \href
  {https://ui.adsabs.harvard.edu/abs/2007ApJ...666.1024B} {666, 1024}

\bibitem[\protect\citeauthoryear{{Boone}}{{Boone}}{2019}]{boone2019}
{Boone} K.,  2019, \mn@doi [\aj] {10.3847/1538-3881/ab5182}, \href
  {https://ui.adsabs.harvard.edu/abs/2019AJ....158..257B} {158, 257}

\bibitem[\protect\citeauthoryear{{Briel}, {Eldridge}, {Stanway}, {Stevance}  \&
  {Chrimes}}{{Briel} et~al.}{2022}]{briel2022}
{Briel} M.~M.,  {Eldridge} J.~J.,  {Stanway} E.~R.,  {Stevance} H.~F.,
  {Chrimes} A.~A.,  2022, \mn@doi [\mnras] {10.1093/mnras/stac1100}, \href
  {https://ui.adsabs.harvard.edu/abs/2022MNRAS.514.1315B} {514, 1315}

\bibitem[\protect\citeauthoryear{{Brown}, {Breeveld}, {Holland}, {Kuin}  \&
  {Pritchard}}{{Brown} et~al.}{2014}]{brown2014}
{Brown} P.~J.,  {Breeveld} A.~A.,  {Holland} S.,  {Kuin} P.,   {Pritchard} T.,
  2014, \mn@doi [\apss] {10.1007/s10509-014-2059-8}, \href
  {https://ui.adsabs.harvard.edu/abs/2014Ap&SS.354...89B} {354, 89}

\bibitem[\protect\citeauthoryear{{Claeys}, {de Mink}, {Pols}, {Eldridge}  \&
  {Baes}}{{Claeys} et~al.}{2011}]{clayes2011}
{Claeys} J.~S.~W.,  {de Mink} S.~E.,  {Pols} O.~R.,  {Eldridge} J.~J.,   {Baes}
  M.,  2011, \mn@doi [\aap] {10.1051/0004-6361/201015410}, \href
  {https://ui.adsabs.harvard.edu/abs/2011A&A...528A.131C} {528, A131}

\bibitem[\protect\citeauthoryear{{Dall'Ora} et~al.,}{{Dall'Ora}
  et~al.}{2014}]{dallora2014}
{Dall'Ora} M.,  et~al., 2014, \mn@doi [\apj] {10.1088/0004-637X/787/2/139},
  \href {https://ui.adsabs.harvard.edu/abs/2014ApJ...787..139D} {787, 139}

\bibitem[\protect\citeauthoryear{{Davis} et~al.,}{{Davis}
  et~al.}{2021}]{davis2021}
{Davis} S.,  et~al., 2021, \mn@doi [\apj] {10.3847/1538-4357/abdd36}, \href
  {https://ui.adsabs.harvard.edu/abs/2021ApJ...909..145D} {909, 145}

\bibitem[\protect\citeauthoryear{{Dessart}, {Hillier}, {Li}  \&
  {Woosley}}{{Dessart} et~al.}{2012}]{dessart12}
{Dessart} L.,  {Hillier} D.~J.,  {Li} C.,   {Woosley} S.,  2012, \mn@doi
  [\mnras] {10.1111/j.1365-2966.2012.21374.x}, \href
  {https://ui.adsabs.harvard.edu/abs/2012MNRAS.424.2139D} {424, 2139}

\bibitem[\protect\citeauthoryear{{Drout} et~al.,}{{Drout}
  et~al.}{2011}]{drout2011}
{Drout} M.~R.,  et~al., 2011, \mn@doi [\apj] {10.1088/0004-637X/741/2/97},
  \href {https://ui.adsabs.harvard.edu/abs/2011ApJ...741...97D} {741, 97}

\bibitem[\protect\citeauthoryear{Duvenaud}{Duvenaud}{2014}]{duvenaud2014}
Duvenaud D.,  2014, PhD thesis, University of Cambridge

\bibitem[\protect\citeauthoryear{{Eldridge} \& {Xiao}}{{Eldridge} \&
  {Xiao}}{2019}]{eldridge2019}
{Eldridge} J.~J.,  {Xiao} L.,  2019, \mn@doi [\mnras] {10.1093/mnrasl/slz030},
  \href {https://ui.adsabs.harvard.edu/abs/2019MNRAS.485L..58E} {485, L58}

\bibitem[\protect\citeauthoryear{{Eldridge}, {Stanway}, {Xiao}, {McClelland},
  {Taylor}, {Ng}, {Greis}  \& {Bray}}{{Eldridge} et~al.}{2017}]{eldridge2017}
{Eldridge} J.~J.,  {Stanway} E.~R.,  {Xiao} L.,  {McClelland} L.~A.~S.,
  {Taylor} G.,  {Ng} M.,  {Greis} S.~M.~L.,   {Bray} J.~C.,  2017, \mn@doi
  [\pasa] {10.1017/pasa.2017.51}, \href
  {https://ui.adsabs.harvard.edu/abs/2017PASA...34...58E} {34, e058}

\bibitem[\protect\citeauthoryear{{Elmhamdi} et~al.,}{{Elmhamdi}
  et~al.}{2003}]{elmhamdi2003}
{Elmhamdi} A.,  et~al., 2003, \mn@doi [\mnras]
  {10.1046/j.1365-8711.2003.06150.x}, \href
  {https://ui.adsabs.harvard.edu/abs/2003MNRAS.338..939E} {338, 939}

\bibitem[\protect\citeauthoryear{{Fakhouri} et~al.,}{{Fakhouri}
  et~al.}{2015}]{fakhouri2015}
{Fakhouri} H.~K.,  et~al., 2015, \mn@doi [\apj] {10.1088/0004-637X/815/1/58},
  \href {https://ui.adsabs.harvard.edu/abs/2015ApJ...815...58F} {815, 58}

\bibitem[\protect\citeauthoryear{{Faran} et~al.,}{{Faran}
  et~al.}{2014}]{faran2014}
{Faran} T.,  et~al., 2014, \mn@doi [\mnras] {10.1093/mnras/stu955}, \href
  {https://ui.adsabs.harvard.edu/abs/2014MNRAS.442..844F} {442, 844}

\bibitem[\protect\citeauthoryear{{Filippenko}}{{Filippenko}}{1988}]{filipenko1988}
{Filippenko} A.~V.,  1988, \mn@doi [\aj] {10.1086/114940}, \href
  {https://ui.adsabs.harvard.edu/abs/1988AJ.....96.1941F} {96, 1941}

\bibitem[\protect\citeauthoryear{{Filippenko}}{{Filippenko}}{1997}]{filippenko1997}
{Filippenko} A.~V.,  1997, \mn@doi [\araa] {10.1146/annurev.astro.35.1.309},
  \href {https://ui.adsabs.harvard.edu/abs/1997ARA&A..35..309F} {35, 309}

\bibitem[\protect\citeauthoryear{{Galbany} et~al.,}{{Galbany}
  et~al.}{2016}]{galbany2016}
{Galbany} L.,  et~al., 2016, \mn@doi [\aj] {10.3847/0004-6256/151/2/33}, \href
  {https://ui.adsabs.harvard.edu/abs/2016AJ....151...33G} {151, 33}

\bibitem[\protect\citeauthoryear{{Ghodla}, {van Zeist}, {Eldridge}, {Stevance}
  \& {Stanway}}{{Ghodla} et~al.}{2022}]{ghodla2022}
{Ghodla} S.,  {van Zeist} W. G.~J.,  {Eldridge} J.~J.,  {Stevance} H.~F.,
  {Stanway} E.~R.,  2022, \mn@doi [\mnras] {10.1093/mnras/stac120}, \href
  {https://ui.adsabs.harvard.edu/abs/2022MNRAS.511.1201G} {511, 1201}

\bibitem[\protect\citeauthoryear{{Guillochon}, {Parrent}, {Kelley}  \&
  {Margutti}}{{Guillochon} et~al.}{2017}]{guillochon2017}
{Guillochon} J.,  {Parrent} J.,  {Kelley} L.~Z.,   {Margutti} R.,  2017,
  \mn@doi [\apj] {10.3847/1538-4357/835/1/64}, \href
  {https://ui.adsabs.harvard.edu/abs/2017ApJ...835...64G} {835, 64}

\bibitem[\protect\citeauthoryear{{Guti{\'e}rrez} et~al.,}{{Guti{\'e}rrez}
  et~al.}{2017}]{gutierrez2017}
{Guti{\'e}rrez} C.~P.,  et~al., 2017, \mn@doi [\apj]
  {10.3847/1538-4357/aa8f52}, \href
  {https://ui.adsabs.harvard.edu/abs/2017ApJ...850...89G} {850, 89}

\bibitem[\protect\citeauthoryear{{Hicken} et~al.,}{{Hicken}
  et~al.}{2017}]{hicken2017}
{Hicken} M.,  et~al., 2017, \mn@doi [\apjs] {10.3847/1538-4365/aa8ef4}, \href
  {https://ui.adsabs.harvard.edu/abs/2017ApJS..233....6H} {233, 6}

\bibitem[\protect\citeauthoryear{{Huang} et~al.,}{{Huang}
  et~al.}{2015}]{huang2015}
{Huang} F.,  et~al., 2015, \mn@doi [\apj] {10.1088/0004-637X/807/1/59}, \href
  {https://ui.adsabs.harvard.edu/abs/2015ApJ...807...59H} {807, 59}

\bibitem[\protect\citeauthoryear{{Kim} et~al.,}{{Kim} et~al.}{2013}]{kim2013}
{Kim} A.~G.,  et~al., 2013, \mn@doi [\apj] {10.1088/0004-637X/766/2/84}, \href
  {https://ui.adsabs.harvard.edu/abs/2013ApJ...766...84K} {766, 84}

\bibitem[\protect\citeauthoryear{{Leonard} et~al.,}{{Leonard}
  et~al.}{2002}]{leonard2002}
{Leonard} D.~C.,  et~al., 2002, \mn@doi [\aj] {10.1086/343771}, \href
  {https://ui.adsabs.harvard.edu/abs/2002AJ....124.2490L} {124, 2490}

\bibitem[\protect\citeauthoryear{{Massey}, {Neugent}, {Dorn-Wallenstein},
  {Eldridge}, {Stanway}  \& {Levesque}}{{Massey} et~al.}{2021}]{massey2021}
{Massey} P.,  {Neugent} K.~F.,  {Dorn-Wallenstein} T.~Z.,  {Eldridge} J.~J.,
  {Stanway} E.~R.,   {Levesque} E.~M.,  2021, \mn@doi [\apj]
  {10.3847/1538-4357/ac15f5}, \href
  {https://ui.adsabs.harvard.edu/abs/2021ApJ...922..177M} {922, 177}

\bibitem[\protect\citeauthoryear{{Maund} \& {Smartt}}{{Maund} \&
  {Smartt}}{2009}]{maund2009}
{Maund} J.~R.,  {Smartt} S.~J.,  2009, \mn@doi [Science]
  {10.1126/science.1170198}, \href
  {https://ui.adsabs.harvard.edu/abs/2009Sci...324..486M} {324, 486}

\bibitem[\protect\citeauthoryear{McAllister}{McAllister}{2017}]{mcallister2017}
McAllister M.~J.,  2017, Photometric Mass Determinations of Eclipsing
  Cataclysmic Variables, \url {https://etheses.whiterose.ac.uk/19647/}

\bibitem[\protect\citeauthoryear{{Minkowski}}{{Minkowski}}{1941}]{minkowski1941}
{Minkowski} R.,  1941, \mn@doi [\pasp] {10.1086/125315}, \href
  {https://ui.adsabs.harvard.edu/abs/1941PASP...53..224M} {53, 224}

\bibitem[\protect\citeauthoryear{{Morales-Garoffolo}
  et~al.,}{{Morales-Garoffolo} et~al.}{2014}]{morales-Garoffolo2014}
{Morales-Garoffolo} A.,  et~al., 2014, \mn@doi [\mnras]
  {10.1093/mnras/stu1837}, \href
  {https://ui.adsabs.harvard.edu/abs/2014MNRAS.445.1647M} {445, 1647}

\bibitem[\protect\citeauthoryear{{Nomoto}, {Suzuki}, {Shigeyama}, {Kumagai},
  {Yamaoka}  \& {Saio}}{{Nomoto} et~al.}{1993}]{nomoto1993}
{Nomoto} K.,  {Suzuki} T.,  {Shigeyama} T.,  {Kumagai} S.,  {Yamaoka} H.,
  {Saio} H.,  1993, \mn@doi [\nat] {10.1038/364507a0}, \href
  {https://ui.adsabs.harvard.edu/abs/1993Natur.364..507N} {364, 507}

\bibitem[\protect\citeauthoryear{{Okyudo}, {Kato}, {Ishida}, {Tokimasa}  \&
  {Yamaoka}}{{Okyudo} et~al.}{1993}]{okyudo1993}
{Okyudo} M.,  {Kato} T.,  {Ishida} T.,  {Tokimasa} N.,   {Yamaoka} H.,  1993,
  \pasj, \href {https://ui.adsabs.harvard.edu/abs/1993PASJ...45L..63O} {45,
  L63}

\bibitem[\protect\citeauthoryear{Paciorek \& Schervish}{Paciorek \&
  Schervish}{2003}]{paciorek2003}
Paciorek C.,  Schervish M.,  2003, Advances in neural information processing
  systems, 16

\bibitem[\protect\citeauthoryear{{Pastorello} et~al.,}{{Pastorello}
  et~al.}{2008}]{pastorello2008}
{Pastorello} A.,  et~al., 2008, \mn@doi [\mnras]
  {10.1111/j.1365-2966.2008.13618.x}, \href
  {https://ui.adsabs.harvard.edu/abs/2008MNRAS.389..955P} {389, 955}

\bibitem[\protect\citeauthoryear{Pedregosa et~al.,}{Pedregosa
  et~al.}{2011}]{scikit-learn}
Pedregosa F.,  et~al., 2011, Journal of Machine Learning Research, 12, 2825

\bibitem[\protect\citeauthoryear{{Pessi} et~al.,}{{Pessi}
  et~al.}{2019}]{pessi2019}
{Pessi} P.~J.,  et~al., 2019, \mn@doi [\mnras] {10.1093/mnras/stz1855}, \href
  {https://ui.adsabs.harvard.edu/abs/2019MNRAS.488.4239P} {488, 4239}

\bibitem[\protect\citeauthoryear{{Podsiadlowski}, {Joss}  \&
  {Hsu}}{{Podsiadlowski} et~al.}{1992}]{podsiadlowski1992}
{Podsiadlowski} P.,  {Joss} P.~C.,   {Hsu} J.~J.~L.,  1992, \mn@doi [\apj]
  {10.1086/171341}, \href
  {https://ui.adsabs.harvard.edu/abs/1992ApJ...391..246P} {391, 246}

\bibitem[\protect\citeauthoryear{{Podsiadlowski}, {Hsu}, {Joss}  \&
  {Ross}}{{Podsiadlowski} et~al.}{1993}]{podsiadlowski1993}
{Podsiadlowski} P.,  {Hsu} J.~J.~L.,  {Joss} P.~C.,   {Ross} R.~R.,  1993,
  \mn@doi [\nat] {10.1038/364509a0}, \href
  {https://ui.adsabs.harvard.edu/abs/1993Natur.364..509P} {364, 509}

\bibitem[\protect\citeauthoryear{Rasmussen \& Williams}{Rasmussen \&
  Williams}{2006}]{rasmussenw06}
Rasmussen C.~E.,  Williams C. K.~I.,  2006, Gaussian processes for machine
  learning..
Adaptive computation and machine learning, MIT Press

\bibitem[\protect\citeauthoryear{{Richmond}, {Treffers}, {Filippenko}  \&
  {Paik}}{{Richmond} et~al.}{1996}]{richmond1996}
{Richmond} M.~W.,  {Treffers} R.~R.,  {Filippenko} A.~V.,   {Paik} Y.,  1996,
  \mn@doi [\aj] {10.1086/118048}, \href
  {https://ui.adsabs.harvard.edu/abs/1996AJ....112..732R} {112, 732}

\bibitem[\protect\citeauthoryear{{Rosdahl} et~al.,}{{Rosdahl}
  et~al.}{2018}]{rosdahl2018}
{Rosdahl} J.,  et~al., 2018, \mn@doi [\mnras] {10.1093/mnras/sty1655}, \href
  {https://ui.adsabs.harvard.edu/abs/2018MNRAS.479..994R} {479, 994}

\bibitem[\protect\citeauthoryear{{Sanders} et~al.,}{{Sanders}
  et~al.}{2015}]{sanders2015}
{Sanders} N.~E.,  et~al., 2015, \mn@doi [\apj] {10.1088/0004-637X/799/2/208},
  \href {https://ui.adsabs.harvard.edu/abs/2015ApJ...799..208S} {799, 208}

\bibitem[\protect\citeauthoryear{{Smith}, {Li}, {Filippenko}  \&
  {Chornock}}{{Smith} et~al.}{2011}]{smith2011}
{Smith} N.,  {Li} W.,  {Filippenko} A.~V.,   {Chornock} R.,  2011, \mn@doi
  [\mnras] {10.1111/j.1365-2966.2011.17229.x}, \href
  {https://ui.adsabs.harvard.edu/abs/2011MNRAS.412.1522S} {412, 1522}

\bibitem[\protect\citeauthoryear{{Smith} et~al.,}{{Smith}
  et~al.}{2020}]{smith2020}
{Smith} K.~W.,  et~al., 2020, \mn@doi [\pasp] {10.1088/1538-3873/ab936e}, \href
  {https://ui.adsabs.harvard.edu/abs/2020PASP..132h5002S} {132, 085002}

\bibitem[\protect\citeauthoryear{{Stancliffe} \& {Eldridge}}{{Stancliffe} \&
  {Eldridge}}{2009}]{stancliffe2009}
{Stancliffe} R.~J.,  {Eldridge} J.~J.,  2009, \mn@doi [\mnras]
  {10.1111/j.1365-2966.2009.14849.x}, \href
  {https://ui.adsabs.harvard.edu/abs/2009MNRAS.396.1699S} {396, 1699}

\bibitem[\protect\citeauthoryear{{Stanway} \& {Eldridge}}{{Stanway} \&
  {Eldridge}}{2018}]{stanway2018}
{Stanway} E.~R.,  {Eldridge} J.~J.,  2018, \mn@doi [\mnras]
  {10.1093/mnras/sty1353}, \href
  {https://ui.adsabs.harvard.edu/abs/2018MNRAS.479...75S} {479, 75}

\bibitem[\protect\citeauthoryear{{Stanway}, {Chrimes}, {Eldridge}  \&
  {Stevance}}{{Stanway} et~al.}{2020a}]{stanway2020b}
{Stanway} E.~R.,  {Chrimes} A.~A.,  {Eldridge} J.~J.,   {Stevance} H.~F.,
  2020a, \mn@doi [\mnras] {10.1093/mnras/staa1166}, \href
  {https://ui.adsabs.harvard.edu/abs/2020MNRAS.495.4605S} {495, 4605}

\bibitem[\protect\citeauthoryear{{Stanway}, {Eldridge}  \& {Chrimes}}{{Stanway}
  et~al.}{2020b}]{stanway2020a}
{Stanway} E.~R.,  {Eldridge} J.~J.,   {Chrimes} A.~A.,  2020b, \mn@doi [\mnras]
  {10.1093/mnras/staa2089}, \href
  {https://ui.adsabs.harvard.edu/abs/2020MNRAS.497.2201S} {497, 2201}

\bibitem[\protect\citeauthoryear{{Stevance} \& {Eldridge}}{{Stevance} \&
  {Eldridge}}{2021}]{stevance2021}
{Stevance} H.~F.,  {Eldridge} J.~J.,  2021, \mn@doi [\mnras]
  {10.1093/mnrasl/slab039}, \href
  {https://ui.adsabs.harvard.edu/abs/2021MNRAS.504L..51S} {504, L51}

\bibitem[\protect\citeauthoryear{{Stevance} et~al.,}{{Stevance}
  et~al.}{2016}]{stevance2016}
{Stevance} H.~F.,  et~al., 2016, \mn@doi [\mnras] {10.1093/mnras/stw1479},
  \href {https://ui.adsabs.harvard.edu/abs/2016MNRAS.461.2019S} {461, 2019}

\bibitem[\protect\citeauthoryear{{Stevance}, {Eldridge}  \&
  {Stanway}}{{Stevance} et~al.}{2020}]{stevance2020}
{Stevance} H.,  {Eldridge} J.,   {Stanway} E.,  2020, \mn@doi [The Journal of
  Open Source Software] {10.21105/joss.01987}, \href
  {https://ui.adsabs.harvard.edu/abs/2020JOSS....5.1987S} {5, 1987}

\bibitem[\protect\citeauthoryear{{Stevance}, {Parsons}  \&
  {Eldridge}}{{Stevance} et~al.}{2022}]{stevance2022}
{Stevance} H.~F.,  {Parsons} S.~G.,   {Eldridge} J.~J.,  2022, \mn@doi [\mnras]
  {10.1093/mnrasl/slac001}, \href
  {https://ui.adsabs.harvard.edu/abs/2022MNRAS.511L..77S} {511, L77}

\bibitem[\protect\citeauthoryear{{Stritzinger} et~al.,}{{Stritzinger}
  et~al.}{2018}]{stritzinger2018}
{Stritzinger} M.~D.,  et~al., 2018, \mn@doi [\aap]
  {10.1051/0004-6361/201730842}, \href
  {https://ui.adsabs.harvard.edu/abs/2018A&A...609A.134S} {609, A134}

\bibitem[\protect\citeauthoryear{{Tak{\'a}ts} et~al.,}{{Tak{\'a}ts}
  et~al.}{2014}]{takatas2014}
{Tak{\'a}ts} K.,  et~al., 2014, \mn@doi [\mnras] {10.1093/mnras/stt2203}, \href
  {https://ui.adsabs.harvard.edu/abs/2014MNRAS.438..368T} {438, 368}

\bibitem[\protect\citeauthoryear{{Tak{\'a}ts} et~al.,}{{Tak{\'a}ts}
  et~al.}{2015}]{takats2015}
{Tak{\'a}ts} K.,  et~al., 2015, \mn@doi [\mnras] {10.1093/mnras/stv857}, \href
  {https://ui.adsabs.harvard.edu/abs/2015MNRAS.450.3137T} {450, 3137}

\bibitem[\protect\citeauthoryear{{The GPyOpt Authors}}{{The GPyOpt
  Authors}}{2016}]{gpyopt2016}
{The GPyOpt Authors} 2016, GPyOpt: A Bayesian Optimization framework in Python,
  \url{http://github.com/SheffieldML/GPyOpt}

\bibitem[\protect\citeauthoryear{{Tsvetkov}, {Volkov}, {Baklanov}, {Blinnikov}
  \& {Tuchin}}{{Tsvetkov} et~al.}{2009}]{tsvetkov2009}
{Tsvetkov} D.~Y.,  {Volkov} I.~M.,  {Baklanov} P.,  {Blinnikov} S.,   {Tuchin}
  O.,  2009, Peremennye Zvezdy, \href
  {https://ui.adsabs.harvard.edu/abs/2009PZ.....29....2T} {29, 2}

\bibitem[\protect\citeauthoryear{{Valenti} et~al.,}{{Valenti}
  et~al.}{2016}]{valenti2016}
{Valenti} S.,  et~al., 2016, \mn@doi [\mnras] {10.1093/mnras/stw870}, \href
  {https://ui.adsabs.harvard.edu/abs/2016MNRAS.459.3939V} {459, 3939}

\bibitem[\protect\citeauthoryear{{Villar} et~al.,}{{Villar}
  et~al.}{2019}]{villar2019}
{Villar} V.~A.,  et~al., 2019, \mn@doi [\apj] {10.3847/1538-4357/ab418c}, \href
  {https://ui.adsabs.harvard.edu/abs/2019ApJ...884...83V} {884, 83}

\bibitem[\protect\citeauthoryear{{Yoon}, {Dessart}  \& {Clocchiatti}}{{Yoon}
  et~al.}{2017}]{yoon2017}
{Yoon} S.-C.,  {Dessart} L.,   {Clocchiatti} A.,  2017, \mn@doi [\apj]
  {10.3847/1538-4357/aa6afe}, \href
  {https://ui.adsabs.harvard.edu/abs/2017ApJ...840...10Y} {840, 10}

\bibitem[\protect\citeauthoryear{{Young} et~al.,}{{Young}
  et~al.}{2006}]{young2006}
{Young} P.~A.,  et~al., 2006, \mn@doi [\apj] {10.1086/500108}, \href
  {https://ui.adsabs.harvard.edu/abs/2006ApJ...640..891Y} {640, 891}

\bibitem[\protect\citeauthoryear{{Yuan} et~al.,}{{Yuan}
  et~al.}{2016}]{yuan2016}
{Yuan} F.,  et~al., 2016, \mn@doi [\mnras] {10.1093/mnras/stw1419}, \href
  {https://ui.adsabs.harvard.edu/abs/2016MNRAS.461.2003Y} {461, 2003}

\bibitem[\protect\citeauthoryear{{de Jaeger} et~al.,}{{de Jaeger}
  et~al.}{2019}]{deJaeger2019}
{de Jaeger} T.,  et~al., 2019, \mn@doi [\mnras] {10.1093/mnras/stz2714}, \href
  {https://ui.adsabs.harvard.edu/abs/2019MNRAS.490.2799D} {490, 2799}

\bibitem[\protect\citeauthoryear{{van Driel} et~al.,}{{van Driel}
  et~al.}{1993}]{vandriel1993}
{van Driel} W.,  et~al., 1993, \pasj, \href
  {https://ui.adsabs.harvard.edu/abs/1993PASJ...45L..59V} {45, L59}

\makeatother
\end{thebibliography}




\appendix

\section{Sample Light curves}
\begin{figure*}
	\includegraphics[width=15cm]{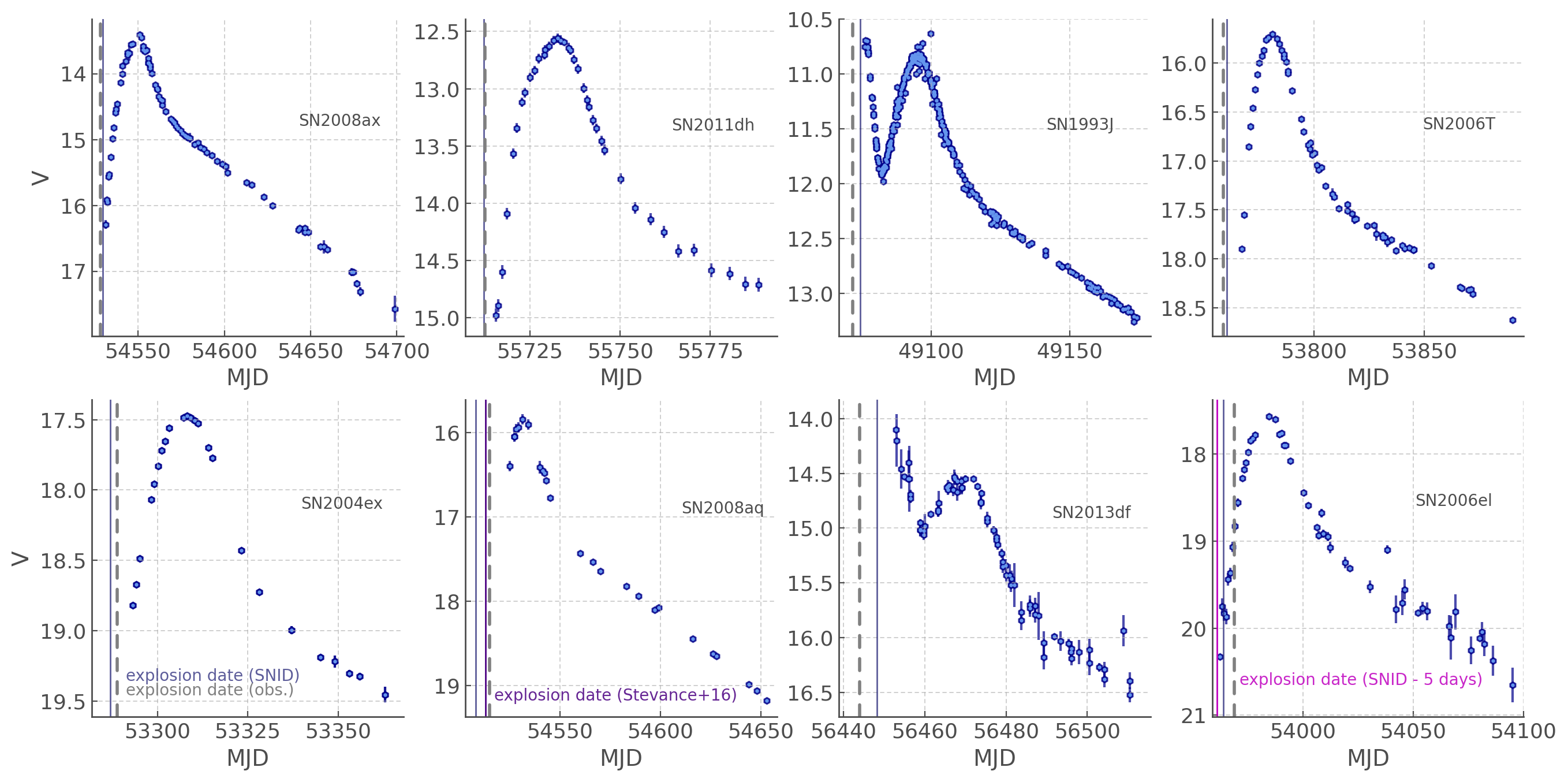}
    \caption{Light curves in our sample of type IIb SNe. Note that we show both the observationally constrained (grey) and the SNID (blue) explosion dates presented in \citetalias{pessi2019} in all light curves to illustrate the similarities (or differences) -- the explosion date we use in practice is shown in Table \ref{tab:sn_sample}. }
    \label{fig:LC_IIb}
\end{figure*}

\begin{figure*}
	\includegraphics[width=15cm]{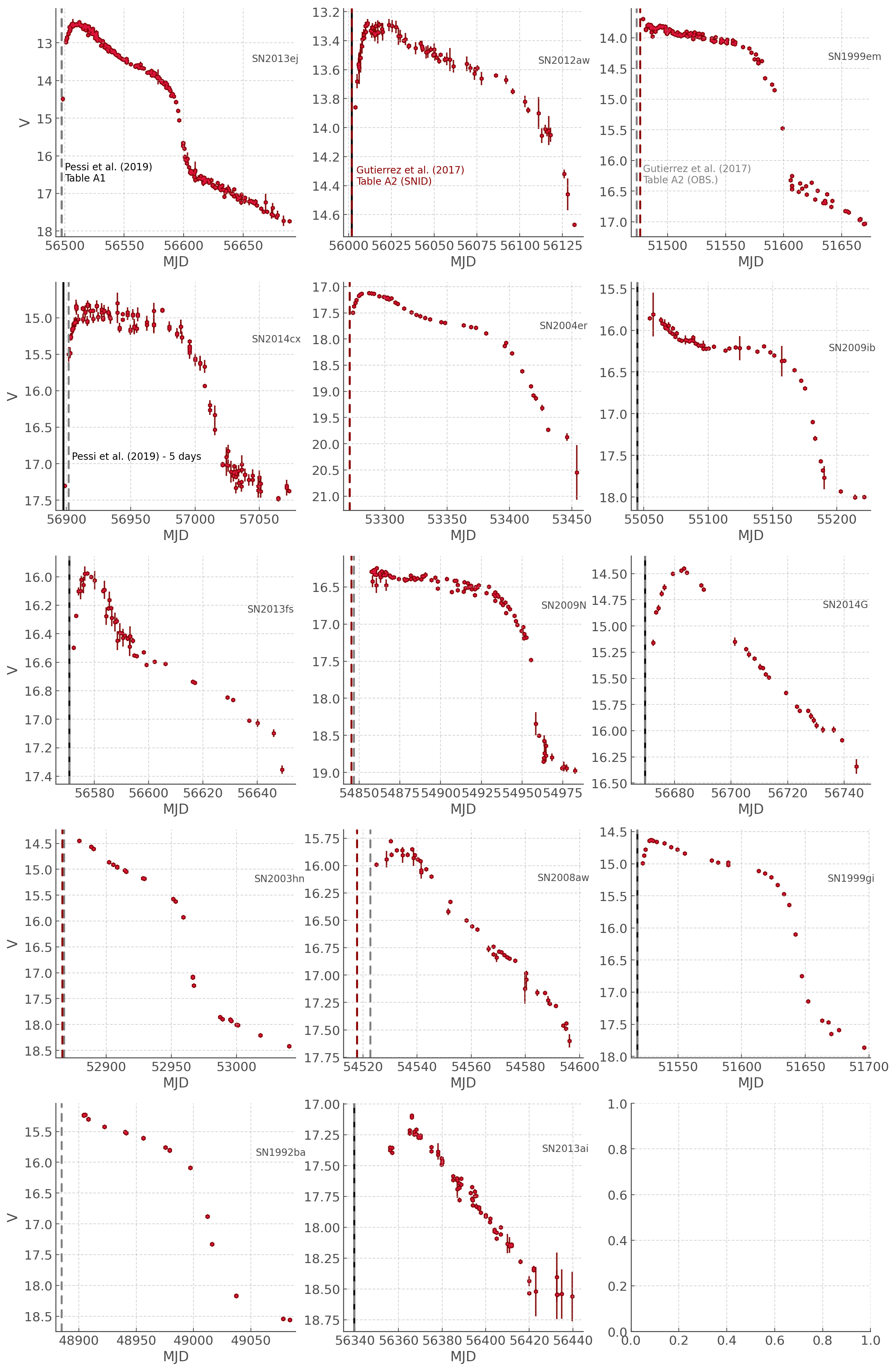}
    \caption{Light curves in our sample of type II SNe. Note that when both the observationally constrained (Grey) and the SNID (Red) explosion dates are reported in the literature they are shown here to illustrate the similarities (or differences) -- the explosion date we use in practice is shown in Table \ref{tab:sn_sample}. The explosion dates used by \citetalias{pessi2019} are shown with a solid black line.}
    \label{fig:LC_II}
\end{figure*}


\bsp	
\label{lastpage}
\end{document}